\documentclass[pre,twocolumn,showpacs]{revtex4-1}
\usepackage{amsmath,graphicx}

\usepackage[colorlinks=true,linkcolor=blue,urlcolor=blue,citecolor=blue]{hyperref}

\usepackage[caption=false]{subfig}

\newcount\minute
\newcount\hour
\newcount\hourMins
\def\now
{
  \minute=\time
  \hour=\time \divide \hour by 60
  \hourMins=\hour \multiply\hourMins by 60
  \advance\minute by -\hourMins
  \zeroPadTwo{\the\hour}:\zeroPadTwo{\the\minute}
}\def\timestamp
{
  \today\ \now
}
\def\today
{
  \zeroPadTwo{\the\day}-\zeroPadTwo{\the\month}-\the\year%
}
\def\zeroPadTwo#1%
{% Left zero pad of the argument to 2 digits.  The argument
  \ifnum #1<10 0\fi
  #1%
}
\date{\timestamp -\jobname.tex}

\def\vep{\varepsilon}

\newcommand{\be}{\begin{equation}}
\newcommand{\ee}{\end{equation}}
\newcommand{\bea}{\begin{eqnarray}}
\newcommand{\eea}{\end{eqnarray}}
\newcommand{\ba}{\begin{array}}
\newcommand{\ea}{\end{array}}

\newcommand{\eal}{\end{align}}

\newcommand{\bem}{\begin{multline}}
\newcommand{\eem}{\end{multline}}
\def\eq#1{Eq.~(\ref{#1})}
\def\la{\langle}
\def\ra{\rangle}

\def\da{\dagger}
\def\kf{{k_F}}
\def\fsk{|\Omega\ra}
\def\fsb{\la \Omega|}
\def\vak{|0\ra}
\def\vab{\la 0|}

\def\pb{\psi}
\def\pf{\Psi}

\begin{document} 

%%%%%%%%%%%%%%%%%%%%%%%%%%%%
\title{Quasiparticle decay in a one-dimensional Bose-Fermi mixture}
\author{Benjamin Reichert}
\author{Aleksandra Petkovi\'{c}}
\author{Zoran Ristivojevic}
\date{\today\now}
\affiliation{Laboratoire de Physique Th\'{e}orique, Universit\'{e} de Toulouse, CNRS, UPS, 31062 Toulouse, France}
%%%%%%%%%%%%%%%%%%%%%%%%%%%%

\begin{abstract}
In a one-dimensional weakly interacting Bose-Fermi mixture one branch of elementary excitations is well described by the Bogoliubov spectrum. Here we use the microscopic theory to study the decay of such quasiparticle excitations. The main scattering process which leads to their decay is the backscattering of a Bogoliubov quasiparticle off the Fermi sea, where a particle-hole pair is excited. For a low-momentum quasiparticle (phonon) of momentum $q$, we find that the decay rate scales as $q^3$ provided $q$ is smaller than the Fermi momentum $k_F$, while in the opposite case the decay behaves as $q^2$. If the ratio of the masses of fermions and bosons is equal to the ratio of the boson-fermion and the boson-boson interaction strengths, the decay rate changes dramatically. It scales as $q^7$ for $q<k_F$, while we find $q^6$ scaling at $q>k_F$. For a high-momentum Bogoliubov quasiparticle, we find a constant decay rate for $q<k_F$, while it scales as $1/q$ for $q>k_F$. We also find an analytic expression for the decay rate in the crossover region between low and high momenta. The decay rate is a continuous, but nonanalytic function of the momentum at $q=k_F$. In the special case when the parameters of our system correspond to the integrable model, we observe that the decay rate vanishes. 
\end{abstract}

\maketitle

\section{Introduction}

For a long time the investigation of mixtures of quantum fluids was limited to the $^3$He--$^4$He mixture. However, highly controlled experimental setups with ultracold atomic gases have been developed in the last 20 years \cite{bloch2008many-body}, which provide a new playground for detailed studies of quantum mixtures. In particular, various experimental realizations of Bose-Fermi mixtures have been achieved \cite{schreck2001quasipure,hadzibabic2002two-species,modugno2002two,goldwin2004measurement,ospelkaus2006interaction-driven,ferrier-barbut2014mixture}. Such realizations with cold gases are especially interesting due to the  possibility to tune the interaction between atoms using the Feshbach resonance technique, which enables one to probe the system at arbitrary coupling. Moreover, one is able to create low-dimensional mixtures by suitably applying the trapping potential and thus study phenomena in reduced dimensions where the role of quantum fluctuations is enhanced. Ground-state properties, thermodynamics, and the rich phase diagram of one-dimensional Bose-Fermi mixtures are studied in numerous works \cite{lai1971ground-state,das_bose-fermi_2003,   cazalilla_instabilities_2003,mathey2004luttinger,takeuchi_mixing-demixing_2005,batchelor_exact_2005,frahm2005correlation,imambekov2006exactly,imambekov_applications_2006,pollet2006phase,adhikari_one-dimensional_2007,mathey_phase_2007,sengupta2007quantum,rizzi_pairing_2008,guan2008magnetic,fang2011exact,yin2012quantum,guan2013fermi,hu2016strongly}.

The usual starting point for the theoretical analysis of one-dimensional interacting bosonic and fermionic systems is the Luttinger liquid theory \cite{cazalilla2011one}. Within this hydrodynamic approach, the low-energy excitations are bosonic quasiparticles with linear energy spectrum. This poses a problem for the calculation of the quasiparticle decay rate \cite{imambekov_one-dimensional_2012}. 
Within the framework of the Luttinger liquid theory, these quasiparticles have an infinite lifetime. However, the quasiparticles are not exact eigenstates of the system and therefore should decay. In order to describe their decay, we have to go beyond the Luttinger liquid description and carefully account for various nonlinearities \cite{imambekov_one-dimensional_2012,lin_thermalization_2013,ristivojevic2016decay}. The decay rate of quasiparticles is a quantity that could be directly probed, e.g., by measuring the dynamic structure factor. The latter does not have the form of an infinitely sharp $\delta$ function at the position of the excitation, but rather a peak with the width determined by the decay rate.

The decay of quasiparticles in one-dimensional quantum liquids has recently attracted considerable attention  \cite{khodas2007fermi-luttinger, gangardt_quantum_2010, tan_relaxation_2010, karzig_energy_2010,
micklitz2011thermalization, ristivojevic2013relaxation,
lin_thermalization_2013, matveev_decay_2013,
ristivojevic2014decay,protopopov2014relaxation,
protopopov_equilibration_2015,ristivojevic2016decay}. For both systems of interacting bosons and fermions, the lowest-energy excitations are fermionic quasiparticles \cite{rozhkov2005fermionic,imambekov_one-dimensional_2012}. Their decay rate behaves as the eighth power of the momentum \cite{matveev_decay_2013}. The nature and the decay rate of quasiparticle excitations at higher momenta become more complicated. For example, in a weakly interacting Bose gas, good quasiparticles are bosonic quasiparticles that are  well described by the Bogoliubov spectrum \cite{kulish1976comparison,imambekov_one-dimensional_2012,ristivojevic2016decay}. At low momenta, the quasiparticles are phonons and their decay rate scales as the seventh power of the momentum \cite{ristivojevic2014decay,ristivojevic2016decay}. At higher momenta, the decay rate of Bogoliubov quasiparticles crosses over into a momentum-independent value \cite{tan_relaxation_2010,ristivojevic2016decay}.

A one-dimensional Bose-Fermi mixture is commonly described in terms of a two-component Luttinger liquid \cite{cazalilla_instabilities_2003,mathey2004luttinger,orignac_competition_2010}. Although this theory is fruitful for some questions, to account for the decay of quasiparticle excitations, a more detailed theoretical description is needed. For particular values of parameters, the system has a special feature of being integrable. This means that one can find the exact solution using the Bethe ansatz method \cite{lai1971ground-state,batchelor_exact_2005,imambekov_applications_2006}. However, in this case, a large number of conservation laws prevent equilibration, and quasiparticles do not decay. The low-energy microscopic description should be consistent and in special cases, which correspond to the integrable models, it must be in agreement with the results obtained by the Bethe ansatz.

In this paper, we study a Bose-Fermi mixture at weak interaction. In this regime, one branch of excitations is inherited from weakly interacting bosons and has a Bogoliubov form. We study the effect of Bose-Fermi coupling on the decay of Bogoliubov excitations. We use the hydrodynamic microscopic description which accounts for various anharmonicities in the theory. We consider spin-polarized fermions. At zero temperature, the leading process that gives the decay is the quasiparticle backscattering off the Fermi sea, where a particle-hole pair is created. For generic values of the coupling constants and the masses of bosons and fermions, we find different regimes of scaling of the decay rate. These regimes depend on the initial quasiparticle momentum, Fermi wave vector, and the characteristic momentum of Bogoliubov excitations. However, we show that by tuning the ratio of the masses of bosons and fermions and the interactions, different regimes emerge where the decay rate dramatically changes.

This paper is organized as follows. In Sec.~\ref{sec:model}, we introduce the model for a one-dimensional Bose-Fermi mixture. In Sec.~\ref{sec:amp}, we calculate the scattering matrix element describing the decay of a Bogoliubov quasiparticle using the perturbation theory in the interaction strength. We also analyze the matrix element in the limits of small and large momenta. 
In Sec.~\ref{sec:dr}, we evaluate the decay rate of a Bogoliubov quasiparticle. We consider regions of low- and high-energy excitations, and provide the analytic expression for the full crossover function between them. 
Section~\ref{sec:stability} contains discussions and conclusions. The technical details of the calculations are given in Appendices \ref{app:amp}--\ref{app:eqcorr}. 

\section{Microscopic model}
\label{sec:model}

We study a weakly interacting Bose-Fermi mixture described by the Hamiltonian $H=H_B+H_F+V_1$, where
	\begin{gather}
		{H_B=\dfrac{\hbar^2}{2m}\int dx \,( \nabla \pb^\da)    ( \nabla \pb)    +\frac{g}{2}  \int dx  \,n^2 , 	\label{eq:Hb}}\\
		{H_F= \dfrac{\hbar^2}{2M} \int dx  \,( \nabla \pf^\da)( \nabla \pf ) , 	\label{eq:Hf}}\\		
		{V_1=  G \int dx\, n \,n_F.	\label{eq:V1}}
	\end{gather}
Equation (\ref{eq:Hb}) describes bosonic particles of the mass $m$ that interact via contact repulsion of the strength $g$. By $\psi$ and $\psi^\dagger$, we denote bosonic single-particle field operators that satisfy the standard commutation relation $[\psi(x),\psi^\dagger(y)]=\delta(x-y)$. The density of bosons is $n=\psi^\dagger\psi$. Equation (\ref{eq:Hf}) describes noninteracting spinless fermionic particles of the mass $M$. By $\Psi$ and $\Psi^\dagger$, we denote fermionic single particle field operators that satisfy the standard anticommutation relation $\{\Psi(x),\Psi^\dagger(y)\}=\delta(x-y)$. Equation (\ref{eq:V1}) describes the mutual interaction between two subsystems, where $G$ is the interaction strength, while $n_F=\Psi^\dagger \Psi$ denotes the density of fermions. 

We note that in the special case  $m=M$ and $g=G$, the Hamiltonian $H$ describes an integrable model which can be solved using the Bethe ansatz technique \cite{lai1971ground-state,batchelor_exact_2005,imambekov_applications_2006}. We also note that in the case of fermions with spin, the integrable model is realized under the conditions $m=M$ and $g=G$, but with an additional requirement of contact interaction between fermions of the strength $g$ \cite{lai1971ground-state,batchelor_exact_2005,imambekov_applications_2006}.

Our goal is to find the excitations of the system as well as small residual interaction between them. We begin with the subsystem of bosons that is described by the Hamiltonian (\ref{eq:Hb}). In the weakly interacting case, we use a hydrodynamic approach where the single-particle operator is expressed as \cite{popov1972theory,haldane_effective_1981}
\be\label{madelung}
		\pb^\da(x)=\sqrt{n(x)}\,e^{i\theta(x)}.
\ee
Here the density $n$ and the phase $\theta$ obey the bosonic commutation relation $[n(x),\theta(y)]=-i\delta(x-y)$.
The  Hamiltonian (\ref{eq:Hb}) then becomes
	\be
		H_B=\frac{\hbar^2}{2m} \int dx\left[ n (\nabla \theta)^2+\frac{(\nabla n)^2}{4n}	\right] +\frac{g}{2}  \int dx  \,n^2 .
	\label{eq:hb_nt}
	\ee
To account for the low-energy excitations, we consider small density fluctuations,
	\be \label{density}
		n=n_0+\frac{\nabla \phi}{\pi},
	\ee
around the mean density $n_0$. Here the new field $\phi$ is defined by the commutation relation $[\nabla \phi(x),\theta(y)]=-i\pi\delta(x-y)$. Such hydrodynamic description is valid as long as the density fluctuations are small,  $|\nabla\phi|\ll n_0$, such that the root in Eq.~(\ref{madelung}) stays positive. This occurs at the wave vectors of excitations below $n_0$. We eventually expand the Hamiltonian (\ref{eq:hb_nt}) in small $\nabla \phi/n_0$. The obtained result can be conveniently split in different powers of  $\nabla\theta$ and $\nabla\phi$ as $H_B=H_{B0}+V_3+V_4+\ldots$, where $H_{B0}$ is a quadratic term, $V_{3}$ is a cubic term, $V_4$ is a quartic term, etc.  

The quadratic term  is given by
	\be
		H_{B0}= \frac{\hbar}{2\pi}\int{ dx \left\{  v K\left[ (\nabla \theta)^2+\frac{(\nabla^2 \phi)^2}{4\pi^2n_0^2}\right]   + \dfrac{v}{K}{(\nabla \phi)^2}   \right\} },
		\label{eq:hbo}
	\ee
which corresponds to the usual Luttinger liquid Hamiltonian, but with the additional term  $\propto(\nabla^2\phi)^2$ identified as the quantum pressure.  The sound velocity  $v$ and the Luttinger liquid parameter $K$ are given by
	\be
		v=\sqrt{\dfrac{g n_0}{m}},\quad K=\dfrac{\pi\hbar n_0}{m v}.
		\label{eq:vK}
	\ee
We consider the weakly interacting limit, $K\gg 1$. 
The cubic term $V_3$ reads
	\be
		V_{3}=\frac{\hbar^2}{2\pi m}\int{ dx \left[ (\nabla \phi)(\nabla \theta)^2-\frac{1}{4\pi^2 n_0^2}(\nabla \phi)(\nabla^2\phi)^2         \right] }.
		\label{eq:v3}
	\ee
For the purpose of the scattering process considered in the next section, we do not need the quartic anharmonic term.

 The quadratic term (\ref{eq:hbo}) is diagonalized by expanding the bosonic field $\phi$ and $\theta$ in normal modes as
	\begin{gather}
		\nabla \phi(x)=\sum_q \sqrt{\dfrac{ \pi^2n_0 }{2Lm\vep_q}}|q|e^{iqx/\hbar}(b^\da_{-q}+b_q),\label{eq:nmb1}\\
		\nabla \theta(x)=\sum_q \sqrt{\dfrac{ m\vep_q }{2L\hbar^2 n_0}}\text{sgn}(q)e^{iqx/\hbar}(b^\da_{-q}-b_q).
		\label{eq:nmb2}
	\end{gather}
Here, $L$ is the size of the system, while $b_q^\dagger$ and $b_q$ are, respectively, the bosonic creation and annihilation operators that satisfy the commutation relation $[b_q,b_{q'}^\dagger]=\delta_{q,q'}$. The quadratic part (\ref{eq:hbo}) in the normal mode representation becomes
	\be
		H_{B0}=\sum_q \vep_q b^\da_q b_q,\quad\text{with}\quad\vep_q=v|q|\sqrt{1+\frac{2q^2}{q_0^2}}.	\label{eq:eb}
	\ee
Here, $q_0=\sqrt{8}m v$. Equation (\ref{eq:eb}) describes bosonic quasiparticle excitations characterized by the nonlinear Bogoliubov dispersion. This nonlinearity arises due to the quantum pressure $\propto (\nabla^2\phi)^2$ in \eq{eq:hbo}, which usually does not appear in the Luttinger liquid Hamiltonian. 

The cubic term  (\ref{eq:v3}) describes the interaction between Bogoliubov quasiparticles. It is given by
\begin{align}
V_3={}& \sum_{q_1,q_2,q_3} \,\frac{|q_1q_2q_3|}{\sqrt{\vep_{q_1}\vep_{q_2}\vep_{q_3}}}\Bigl[f_+(q_1,q_2,q_3) \, (b^\da_{q_3} b^\da_{q_2} b^\da_{q_1}+\text{H.c.}) \notag\\
&+ 3f_-(q_1,q_2,q_3) \, (b^\da_{q_3} b^\da_{q_2} b_{-q_1}  +\text{H.c.}) \Bigr]\delta_{q_1+q_2+q_3,0},
\label{eq:V3exp}
\end{align}
where
	\begin{align}
		f_\pm(q_1,q_2,q_3)={}&\dfrac{v^2}{12\sqrt{ 2Lmn_0}}\biggl[\frac{1}{v^2}\left( \frac{\vep_{q_3}\vep_{q_2}}{q_3q_2}  \pm \frac{\vep_{q_1}\vep_{q_2}}{q_1q_2}  \pm\frac{\vep_{q_1}\vep_{q_3}}{q_1q_3} \right)\notag\\ &-\frac{1}{q_0^2}\left( q_1^2+q_2^2+q_3^2\right) \biggr].
	\end{align}

We now  treat the part of the total Hamiltonian that involves fermions, given by Eqs.~(\ref{eq:Hf})  and (\ref{eq:V1}). The former describes the free fermions, while the latter represents the interaction between bosons and fermions. We expand the fermionic field using the normal modes as
	\be
		\pf(x)=\dfrac{1}{\sqrt{L}}\sum_q e^{iqx/\hbar}a_q,\quad \pf^\da(x)=\dfrac{1}{\sqrt{L}}\sum_q e^{-iqx/\hbar}a^\da_q,
		\label{eq:nmf}
	\ee
where the fermion operators satisfy the anticommutation relation
$\{a_q,a_{q'}^\da\}=\delta_{q,q'}$.
The fermions are described by the quadratic Hamiltonian
	\be
		H_{F}=\sum_q E_q a^\da_q a_q, \quad \text{with} \quad E_q=\frac{q^2}{2M}.	\label{eq:ef}
	\ee
In the regime of weak Bose-Fermi coupling $G$, the two branches of excitation of the Bose-Fermi mixture are given by the expressions (\ref{eq:eb}) and (\ref{eq:ef}).

Using the normal mode expansions  (\ref{eq:nmb1}) and (\ref{eq:nmb2}), and \eq{eq:nmf}, the Bose-Fermi interaction (\ref{eq:V1}) becomes 
	\be
		 V_{1}=\sum_{q_1,q_2,q_3} \Gamma (q_3) a^\da_{-q_2}a_{q_1}\left(b^\da_{-q_3}+b_{q_3}\right)\delta_{q_1+q_2+q_3,0} ,
		\label{eq:V1exp}
	\ee
where 
	\be
		 \Gamma(q) = G\sqrt{\dfrac{n_0q^2}{2L m \vep_q}}.
	\ee
In normal modes, Eq.~(\ref{eq:V1exp}) describes the process where a fermion either emits or absorbs one Bogoliubov quasiparticle. We consider the case of weak Bose-Fermi coupling, $G\ll g\sqrt{K}$ (see Appendix \ref{app:eqcorr}).

\section{Scattering matrix element}
\label{sec:amp}

As shown in the previous section, one type of excitations of a weakly interacting Bose-Fermi mixture is Bogoliubov quasiparticles, which have the spectrum given by Eq.~(\ref{eq:eb}). Due to weak residual interactions, these excitations are not exact eigenstates of the full Hamiltonian. Therefore, in general, they do decay. The goal of this paper is to study their decay rate. Residual interaction between bosons described by $V_3$ (and by other terms contained in $H_B$, such as $V_4$) are one possible decay channel for a Bogoliubov quasiparticle. However, the Hamiltonian $H_B$ given by Eq.~(\ref{eq:Hb}) describes the Lieb-Liniger model \cite{lieb_exact_1963}. It is integrable and therefore its excitations do not decay. This has been recently explicitly shown in Ref.~\cite{ristivojevic2016decay}. We should therefore study another decay channel due to interaction with fermions, which is described by $V_1$ (and its combination with residual interaction between bosons that may arise in higher orders of perturbation theory). In this section, we use perturbation theory to calculate the scattering matrix element for the decay of a Bogoliubov quasiparticle due to interaction with fermions. 

\begin{figure}
		\includegraphics[scale=0.4]{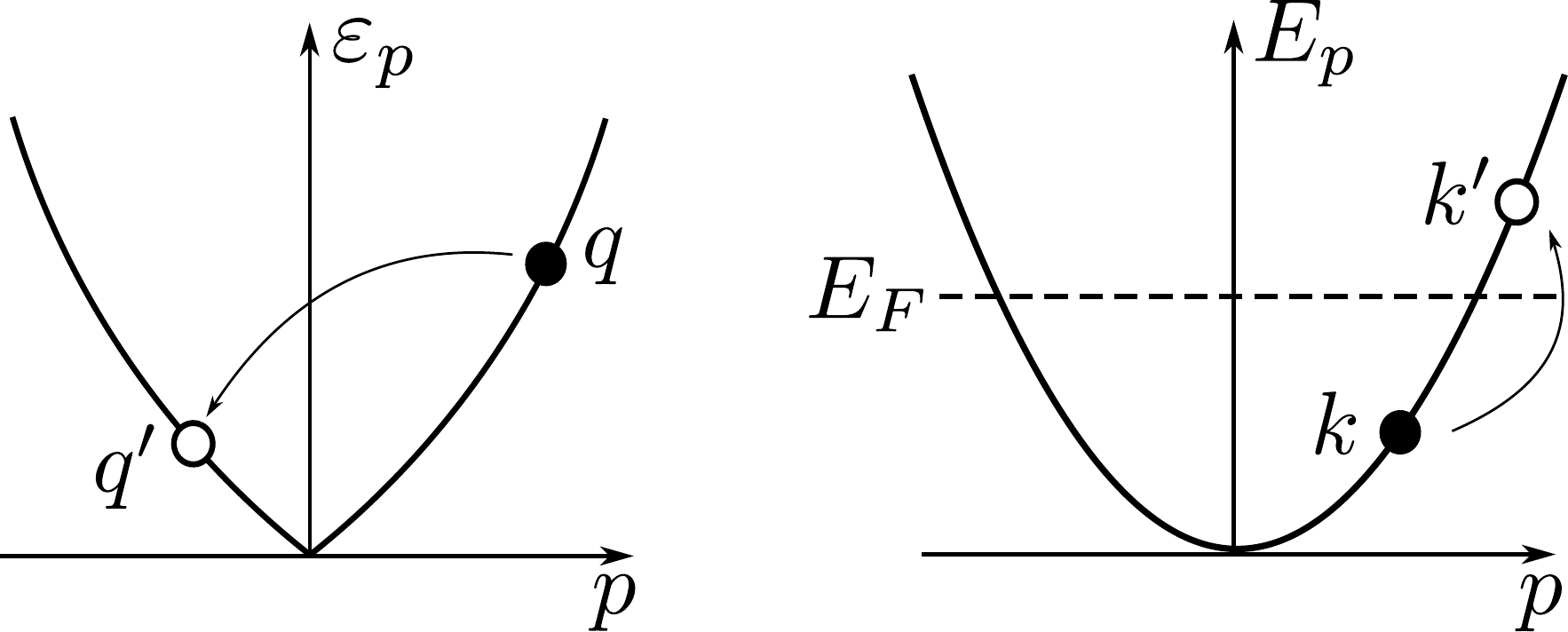}
		\caption{Representation of the decay of a Bogoliubov quasiparticle. Left: The Bogoliubov spectrum of quasiparticles. Right: The spectrum of fermions, where the dashed line represents the Fermi energy.}
		\label{fig:process}
\end{figure}

Using the conservation laws of momentum and energy, one finds that the leading process for the decay of the Bogoliubov quasiparticle is its backscattering off the Fermi sea, where a particle-hole pair becomes excited (see Fig.~\ref{fig:process}). We consider slow fermions, with velocity smaller than the sound velocity $v$. This can be achieved only for sufficiently small Fermi energy. Moreover, we also require that the initial Bogoliubov quasiparticle is of sufficiently small momentum, such that it cannot excite the fermions at too high momenta. If one calculates the correction to the spectrum $E_k$ of fermions due to $V_1$ perturbation (see Appendix \ref{app:Ekcorr}), one finds that it becomes significant only in the very near vicinity of $Mv$, which signals that the quadratic form of the spectrum $E_k=k^2/2M$ is not good only at such high momenta. Therefore, if we consider fermions at momenta below $Mv$, we could safely use the bare fermionic spectrum. On the other hand, fermions cannot be excited at too high momenta if we consider the initial Bogoliubov quasiparticle at sufficiently small momenta, below $Mv/2$. The latter condition is obtained and discussed later in this section.   

The matrix element for the process shown in Fig.~\ref{fig:process} is the central object of our interest. It is given by the expression $\mathcal A_{q,k}^{q',k'}=\vab a_{k'}b_{q'}|T|a_{k}^\da b_q^\da \vak$ in terms of the $T$ matrix. Here, $q$ is the initial momentum of the Bogoliubov quasiparticle, while $q'$ is its final momentum. By $|k|<k_F$ we denote the fermion in the initial state, which is scattered to the state $|k'|>\kf$ above the Fermi sea. By $k_F$, we denote the Fermi momentum, while $|0\rangle$ denotes the vacuum.

For the scattering process that we study, the $T$ matrix is given by the expression
	\be
		T=V+V\dfrac{1}{\vep_q+E_{k}-H_{B0}-H_F}T,
		\label{eq:Tmat}
	\ee
where $V=V_1+V_3$. When the perturbation is weak, \eq{eq:Tmat} can be solved iteratively. Since $V$ does not conserve the number of quasiparticles, the leading contribution to the matrix element is given in second-order perturbation theory in $V$. Therefore, the leading order contribution to the matrix element is
	\be
		\mathcal A_{q,k}^{q',k'}=\sum_m \dfrac{\vab a_{k'}b_{q'}|V|m\ra \la m|V|a_{k}^\da b_q^\da \vak}{\vep_q+E_{k}-E_m},	\label{eq:Ao2}
	\ee
where $E_m=\la m|H_{B0}+H_F|m\ra$ denotes the energy of the intermediate state $|m\ra$. 

Equation (\ref{eq:Ao2}) is evaluated on the mass shell, i.e., accounting for the conservation laws of momentum and energy:
\begin{gather}\label{conservationlaws}
q+k=q'+k',\quad
\varepsilon_q+E_k=\varepsilon_{q'}+E_{k'}.
\end{gather} 
 The details are given in Appendix \ref{app:amp}, while here we give the final result,
\begin{widetext}
\begin{align}
\mathcal A_{q,k}^{q',k'}={}&-\biggl\{\Gamma(q)\Gamma(q')\left(\dfrac{1}{E_{q+k}-E_{k}-\vep_q}+ \dfrac{1}{E_{{q'}-k}-E_{k}+\vep_{q'}} \right) \notag\\
&+ 6 \Gamma(q-q')\dfrac{|q'q  (q-q')|}{\sqrt{\vep_q\vep_{q'}\vep_{q-q'}}}\left[  \dfrac{f_-(-{q'},q,{q'}-q)}{\vep_{q-{q'}}+\vep_q-\vep_{q'}}+\dfrac{f_-(-q,{q'},q-{q'})}{\vep_{q-{q'}}+\vep_{q'}-\vep_q} \right] \biggr\}\delta_{q+k,{q'}+k'}.
		\label{eq:Afinal}
\end{align}
\end{widetext}
It is important to note that the matrix element (\ref{eq:Afinal}) has two contributions, both of them being equally important in the general case. One arises from second-order perturbation theory in $V_1$, while the other is obtained in the same order, but involves both perturbations, $V_1$ and $V_3$. We have verified that in the integrable case ($m=M$ and $g=G$), the matrix element (\ref{eq:Afinal}) nullifies at arbitrary momenta.

One could argue that at second order in $V$, the Bogoliubov quasiparticle could decay via different scattering processes. One possibility is the scattering process where two particle-hole pairs become excited. Another possibility is to obtain three  Bogoliubov quasiparticles in the final state and one particle-hole pair. However, we have verified that on the mass shell, the matrix element for these processes vanishes.

Using the conservation laws (\ref{conservationlaws}), we can express $k$ and $k'$ as functions of $q$ and $q'$:
\begin{align} \label{k}
		k(q,q')&=M\dfrac{\vep_q-\vep_{q'}}{q-q'}-\dfrac{q-q'}{2},\\
		\label{k'}
		k'(q,q')&=M\dfrac{\vep_q-\vep_{q'}}{q-q'}+\dfrac{q-q'}{2}.
\end{align}
We can then reexpress the matrix element (\ref{eq:Afinal}) as 
	\be
	 \mathcal A_{q,k}^{q',k'}=\Lambda(q,q')\delta_{q+k,q'+k'}.
	\label{eq:Akro}
	\ee
The conditions $|k(q,q')|<\kf$ and $|k'(q,q')|>\kf$ impose certain constraints on $q'$ for a given $q$ and $k_F$. We find that $q'$ has to satisfy $q'_{\mathrm{min}}<q'<q'_{\mathrm{max}}$, where the bounds are defined by (see Appendix \ref{q'})
	\begin{align}
		&k(q,q'_\text{max})=-k(q,q'_\text{min})=\kf	&\text{for }\quad\kf<q,	\label{eq:qMm1}
\\
		&k(q,q'_\text{max})=k'(q,q'_\text{min})=\kf	&\text{for }\quad\kf>q.	\label{eq:qMm2}
	\end{align}
For $q'$ outside of the interval $(q'_{\mathrm{min}},q'_{\mathrm{max}})$, the scattering process shown in Fig.~\ref{fig:process} cannot occur, and formally one should define the scattering matrix element to be zero (in the considered order of the perturbation theory) for such region of momenta.

In the above analysis, we have taken into account that $k_F$ is sufficiently low, such that fermions are slow: $|k(q,q')|,|k'(q,q')|<M v$ for all allowed $q'$. This means that we could safely use the bare fermionic spectrum. The condition for $k_F$ is given by 
\begin{align}\label{eq:lowkf}
k_F<Mv-q+q'_*,
\end{align}
where $q'_*$ is defined by the expression $k'(q,q'_*)=M v$ (see Appendix \ref{q'}).

If the inequality (\ref{eq:lowkf}) is satisfied, only the backscattering of the Bogoliubov quasiparticle is allowed. We can assume positive initial momentum $q>0$, and hence $q'<0$. Since $k_F>0$, Eq.~(\ref{eq:lowkf}) implies $Mv-q+q'_*>0$, which imposes a condition on $q$. In the following, we discuss in more detail the two regimes, $q\ll q_0$ and $q \gg q_0$, where we recall $q_0=\sqrt{8}mv$.

%%%%%%%%%%%%%%%%%%%%%%%%%%%%%%%%%%%%%%
\subsection{Small momenta}

We consider the case where the initial Bogoliubov quasiparticle has momentum $q\ll q_0$. This implies also the smallness of the backscattered quasiparticle momentum, $|q'|\ll q_0$. In this regime the Bogoliubov spectrum can be expanded as $\vep_q=v|q|(1+q^2/q_0^2+\ldots)$. For $q>0$,  we expand the scattering matrix element (\ref{eq:Afinal})  at $q'<0$ and obtain
	\be
	\Lambda(q,q')=
	\frac{G}{8 L}\dfrac{(q-q')^2}{m v \sqrt{|qq'|}}     \left(1-\dfrac{G m}{g M}  \right)+\mathcal{O}\left(\dfrac{q^3}{q_0^3} \right).
	\label{eq:Aexp}
	\ee
Here we assumed that the mass of fermions is of the same order or greater than the mass of bosons, $M\gtrsim m$. Note that only the linear part of the Bogoliubov spectrum is necessary to derive Eq.~(\ref{eq:Aexp}). The matrix element (\ref{eq:Aexp}) is valid for momenta $q'_{\mathrm{min}}<q'<q'_{\mathrm{max}}$. Solving Eqs.~(\ref{eq:qMm1}) and (\ref{eq:qMm2}) we find
\be
q'_{\mathrm{min}/\mathrm{max}}=-q\left[1- \dfrac{2\kf}{Mv+\kf} \pm \dfrac{2M^2v^2q}{(Mv+\kf)^3} +\mathcal{O}\left(\dfrac{q^2}{q_0^2} \right) \right] ,
	\label{eq:qmM1}
\ee
for $q<\kf$, and
\be
q'_{\mathrm{min}/\mathrm{max}}=-q\left[1-\dfrac{2(q\mp\kf)}{Mv}+\mathcal{O}\left(\dfrac{q^2}{q_0^2} \right)\right],
	\label{eq:qmM2}
\ee
for $q>\kf$. The Fermi momentum is assumed to satisfy $k_F<M v-\mathcal{O}(q)$, which is obtained by analyzing Eq.~(\ref{eq:lowkf}). We note that in the case $q\to k_F$, one must expand \eq{eq:qmM1} in small $k_F/Mv$. This occurs because $q\ll q_0$ and $M\gtrsim m$ imply $q\ll M v$ and thus $k_F\ll Mv$. After that expansion, one obtains the consistency between $q'_{\text{min}/\text{max}}$ of Eqs.~(\ref{eq:qmM1}) and (\ref{eq:qmM2}). On the other hand, \eq{eq:qmM1} is valid at any ratio $k_F/Mv$.

%%%%%%%%%%%%%%%%%%%%%%%%%%%%%%%%%%%%%%
\subsection{Small momenta, the case $Gm=gM$}

The expanded scattering matrix element (\ref{eq:Aexp}) equals zero for $Gm=gM$. For these particular parameters, there is no reason \textit{a priori} to obtain nullification since it does not correspond to an integrable model. Thus, the scattering matrix element (\ref{eq:Afinal}) should not nullify. Indeed, for $Gm=gM$, we obtain
	\be
	\Lambda(q,q')=
	\frac{G}{128 L}\dfrac{(q-q')^4}{(m v)^3 \sqrt{|qq'|}}\left(1-\frac{m^2}{M^2}\right)+\mathcal{O}\left(\dfrac{q^5}{q_0^5} \right).
	\label{eq:AexpRr}
	\ee
Here we assumed that the mass of fermions is of similar order to or greater than the mass of bosons, $M\gtrsim m$. We point out that in order to obtain \eq{eq:AexpRr}, one needs to take into account the nonlinear part of the Bogoliubov spectrum, whereas in \eq{eq:Aexp}, only the linear part was needed. However, we have the same constraints on $q'$ and $k_F$ as in the previous subsection. In the integrable case  $M=m$ (implying $G=g$), \eq{eq:AexpRr} nullifies, as it must be the case.
%%%%%%%%%%%%%%%%%%%%%%%%%%

\subsection{Large momenta}\label{CCCC}

\begin{figure}
	\centering
	\includegraphics[scale=0.4]{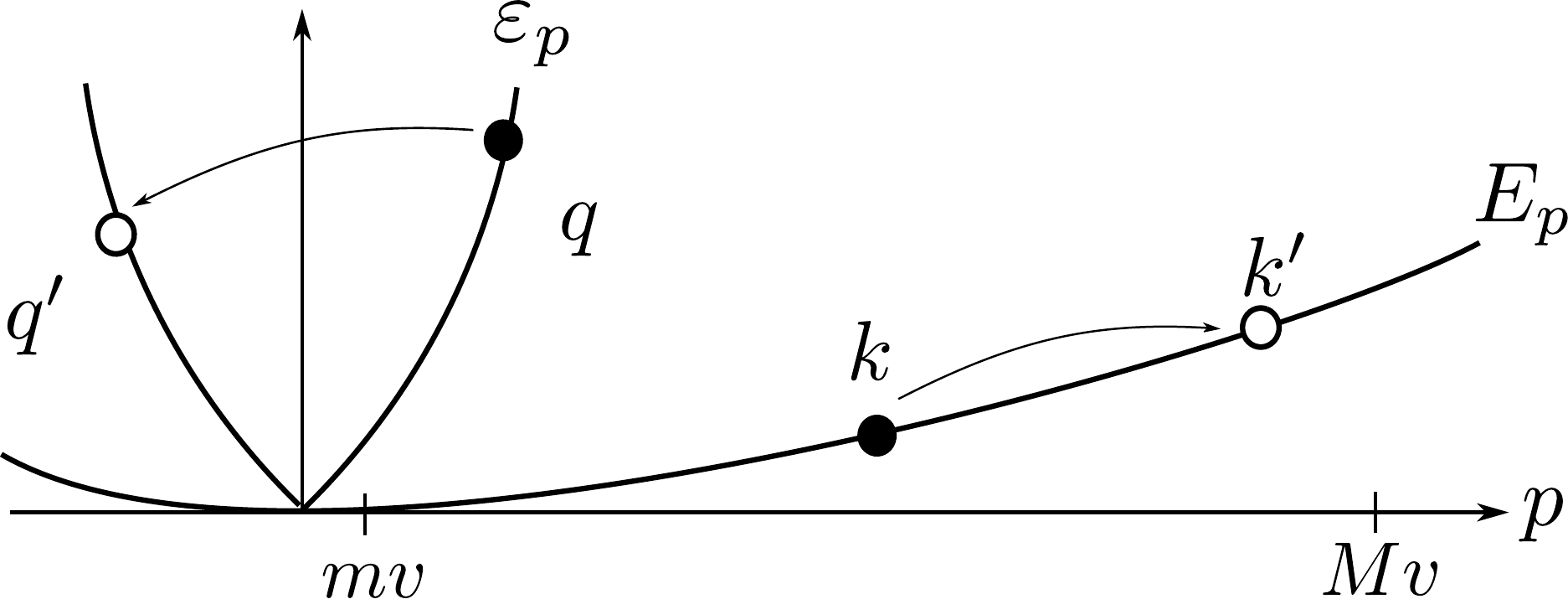}
	\caption{Schematic representation of the decay of a Bogoliubov quasiparticle at momenta $q\gg mv$ and $k,k'<Mv$.}
	\label{fig:process2}
\end{figure}
%
%%%%%%%%%%%%%%%%%%%%%%%%%%%%%

Next we consider the case of high momentum of the Bogoliubov quasiparticle, $q\gg q_0$. At such high momenta, the Bogoliubov spectrum simplifies as $\vep_q\simeq{q}^2/2m$. The scattering matrix element (\ref{eq:Afinal}) becomes
	\be
	\Lambda(q,q')=\frac{G}{L}\left[1+\frac{2m^2v^2}{qq'} \left( 1+\dfrac{2Mm}{M^2-m^2}\dfrac{G}{g} \right)\right]+\mathcal{O}\left(\dfrac{q_0^4}{q^4} \right).
	\label{eq:AexpQ}
	\ee
Here, in order to obtain the dependance on $q$ and $q'$, we have kept the first subleading term. This result assumes that $q'$ is bounded by $q'_\text{min/max}$ that read as 
	\be
	q'_{\mathrm{min}/\mathrm{max}}=\left[-q+\frac{2(k_F\mp q)}{M/m\mp 1}\right]\left[ 1 +\mathcal{O}\left(\dfrac{q_0^4}{q^4} \right) \right] 
	\ee
for $q<\kf$, and
	\be
	q'_{\mathrm{min}/\mathrm{max}}=\left[-q+\frac{2(q\mp k_F)}{M/m+1}\right]\left[ 1 +\mathcal{O}\left(\dfrac{q_0^4}{q^4} \right) \right] 
	\ee
for $q>\kf$. 
Since the fermions are slow, in the case of large $q$ momenta the masses have to satisfy $M\gg m$, as illustrated in Fig.~\ref{fig:process2}. Then, analyzing Eq.~(\ref{eq:lowkf}), we find the constraint on the Fermi momentum, $k_F< M v -2q\left[1+\mathcal{O}(m/M)\right]$. Since $k_F$ is positive, the latter inequality implies that the Bogoliubov quasiparticle must be of sufficiently small momentum, $q<Mv/2$.

\section{Decay rate}
\label{sec:dr}

The scattering matrix element calculated in the previous section can be used for analysis of different kinetic phenomena. Here we focus on the decay rate of a Bogoliubov excitation of momentum $q>0$. It is given by the Fermi golden rule expression
	\begin{align}
		\dfrac{1}{\tau}=\dfrac{2\pi}{\hbar}\sum_{q',k,k'\atop{|k|<\kf<|k'|}} |\mathcal A_{q,k}^{q',k'}|^2\delta(\vep_q+E_{k}- \vep_{q'}-E_{k'}),
		\label{eq:decayrate}
	\end{align}
where $\mathcal A_{q,k}^{q',k'}$ is the scattering matrix element (\ref{eq:Afinal}). Using the energy and momentum conservation laws, the momenta of fermions can be expressed in terms of $q$ and $q'$ by Eqs.~(\ref{k}) and (\ref{k'}).  Then the decay rate (\ref{eq:decayrate}) becomes
	\be
		\dfrac{1}{\tau}=\dfrac{M L^2}{2\pi\hbar^3}\!\int_{q'_\text{min}}^{q'_\text{max}} dq'\,\dfrac{|\Lambda(q,q')|^2}{q-q'},
	\label{eq:decayratefinal}
	\ee
where $q'_\text{min/max}$ are determined by Eqs.~(\ref{eq:qMm1}) and (\ref{eq:qMm2}). 

Although the lower bound of integration $q'_{\mathrm{min}}$ is given by different equations for $q>\kf$ and $q<\kf$, the decay rate is a continuous function of $q$. However, it is not a smooth function at $q=\kf$. As a result, at $q=\kf$ there is a nonanalytic behavior of the decay rate which originates from the abrupt change of the Fermi distribution at the Fermi level at zero temperature. Finite temperature will smoothen this nonanalyticity. Since we consider slow fermions, the Fermi momentum is assumed to satisfy Eq.~(\ref{eq:lowkf}). In Appendix \ref{app:tau}, we present another way to calculate the decay rate.

%%%%%%%%%%%%%%%%
\subsection{Small momenta}

Now we evaluate the decay rate for $q\ll q_0$. We use the scattering matrix element (\ref{eq:Aexp}) in the expression for the decay rate (\ref{eq:decayratefinal}) and obtain, at leading order in $q/q_0$,
	\begin{multline}
		\dfrac{1}{\tau}=\frac{M G^2}{4\pi \hbar^3}\left(\dfrac{G}{g}-\dfrac{M}{m} \right)^2\\
		\times
		\begin{cases}
		\dfrac{q^3(Mv)^3}{(Mv-\kf)(Mv+\kf)^5}&\text{if }q<\kf,  \\[10pt]
		\dfrac{\kf\, q^2}{(Mv)^3}&\text{if }q>\kf. 
		\end{cases}\hspace{13pt}
		\label{eq:dr1}
	\end{multline}
Here, $\kf<{{ M v -\mathcal{O}(q)}}$.  The fermion mass satisfies $M\gtrsim m$.

For $\kf\ll q_0$, one can expand \eq{eq:dr1} and demonstrate the continuity  of the decay rate at $q=\kf$.  Since the result (\ref{eq:dr1}) applies for $M\gtrsim m$, the condition $\kf\ll q_0$ implies $\kf\ll Mv$ and, at leading order in $q/q_0$, the expression    ${q^3(Mv)^3}/{(Mv-\kf)(Mv+\kf)^5}\simeq{q^3}/{(Mv)^3}$. However, \eq{eq:dr1} shows the lack of smoothness at $q=\kf$ caused by the change of the lower bound of integration $q'_{\mathrm{min}}$ at $q=\kf$ in \eq{eq:decayratefinal}. As expected, when all  fermions are removed from  the mixture (i.e.,~$\kf=0$), the decay rate vanishes since bosons do not have any fermions to interact with. In the integrable case, $M=m$ and $G=g$, the decay rate (\ref{eq:dr1}) vanishes, as expected.

%%%%%%%%%%%%%%%%%%%%%%%%%%%%%%%%%%
\subsection{Small momenta, the case $Gm=gM$}

For $gM=Gm$, the leading-order term of the scattering matrix element  (\ref{eq:Aexp}) vanishes and one has to take into account the subleading contribution given by \eq{eq:AexpRr}. This drastically changes  the behavior of the decay rate.   The scattering matrix element (\ref{eq:AexpRr})  is inserted in  \eq{eq:decayratefinal}  and one obtains, at leading order,
	\begin{multline}
		\dfrac{1}{\tau}=\dfrac{M G^2}{64\pi \hbar^3}\dfrac{M^2}{m^2}\left(\dfrac{M^2}{m^2}-1 \right)^2\\
		\times
		\begin{cases}
		\dfrac{q^7(Mv)^3}{(Mv-\kf)(Mv+\kf)^9}&\text{if }q<\kf,  \\[10pt]
		\dfrac{\kf\, q^6}{(Mv)^7}&\text{if }q>\kf.	 
		\end{cases}\hspace{13pt}
		\label{eq:dr2}
	\end{multline}
The fermion mass satisfies $M\gtrsim m$. Also, we assume $\kf<{{ M v -\mathcal{O}(q)}}$. 

Comparing the rates (\ref{eq:dr1}) and (\ref{eq:dr2}), we find that \eq{eq:dr2} is the dominant contribution in the decay rate at
\begin{align}\label{condmM}
\left|1-\frac{mG}{Mg}\right|<\frac{1}{4}\left|\frac{M^2}{m^2}-1\right| \left(\frac{q}{Mv+k_F}\right)^2.
\end{align}
Note that for $M/m=G/g=1$, our system is integrable and the decay rate (\ref{eq:dr2}) vanishes, as expected. We finally note that Eq.~(\ref{eq:dr2}) applies at momenta that are not particularly small, $q\gg mv/\sqrt{K}$. In the opposite case of very small momenta, the Bogoliubov spectrum does not correctly describe good quasiparticle excitations of the bosonic subsystem \cite{imambekov_one-dimensional_2012}.

%%%%%%%%%%%%%%%%%%

\subsection{Large momenta}

Next we consider the decay of a Bogoliubov quasiparticle at large  momenta, $q\gg q_0$. Using the scattering matrix element (\ref{eq:AexpQ}) in the decay rate (\ref{eq:decayratefinal}), we obtain
\be
		\dfrac{1}{\tau}=\frac{M G^2}{2\pi \hbar^3}		\begin{cases}
		\ln{\left(\dfrac{M+m}{M-m}\right)} &\text{if }q<\kf, 	\\[10pt]
		\ln{\left(\dfrac{qM+\kf m}{qM-\kf m}\right)}&\text{if }q>\kf.
		\label{eq:dr3}
		\end{cases}
	\ee
This result applies to slow fermions $\kf<{{ M v -2 q\left[1+\mathcal{O}(m/M)\right]}}$. Since $M v/2>q\gg q_0$, this imposes $M\gg m$. The decay rate (\ref{eq:dr3}) is therefore expanded in $m/M$ and, at the leading order, becomes
	\be
		\dfrac{1}{\tau}=\frac{m G^2}{\pi \hbar^3}
		\begin{cases}
		1 &\text{if }q<\kf, 	\\
		\dfrac{\kf}{q} &\text{if }q>\kf.
		\label{eq:dr4}
		\end{cases}
	\ee
\subsection{The crossover regime}

In the previous sections, we have evaluated the decay rate for Bogoliubov quasiparticles at small and large momenta. However, \eq{eq:decayratefinal} contains the decay rate at arbitrary momenta, which covers the full crossover between the two limiting regimes. It is given by the expression
	\be
		\dfrac{1}{\tau}=\frac{M G^2}{2\pi \hbar^3}F\left(\dfrac{q}{2mv}\right),
		\label{eq:drcross}
	\ee
where the dimensionless crossover function $F(X)$ is given  by
	\begin{align} 
	F(X)={}&\int_{x_\text{min}}^{x_\text{max}} dx \dfrac{(X-x)^3}{\epsilon_X\epsilon_x} \Biggl[\dfrac{Xx(1+2Xx)+2\epsilon_X\epsilon_x}{\epsilon_{X-x}^2-(\epsilon_X-\epsilon_x)^2} \notag\\ &+\dfrac{GM}{gm}\dfrac{X^2x^2}{X^2x^2(X-x)^2-\frac{M^2}{m^2}(x\epsilon_X-X\epsilon_x )^2}\Biggr]^2.
	\end{align}
Here, $\epsilon_z=\sqrt{z^2+z^4}$, while
the dimensionless momenta are $X=q/2mv$,  $x=q'/2mv$, and $x_\text{min/max}=q'_\text{min/max}/2mv$. The bounds of the integration are given by Eqs.~(\ref{eq:qMm1}) and (\ref{eq:qMm2}), while we assume that Eq.~(\ref{eq:lowkf}) is satisfied.  

\begin{figure}
\subfloat[$\kf/2mv=0.01$]{  \includegraphics[clip,width=0.9\columnwidth]{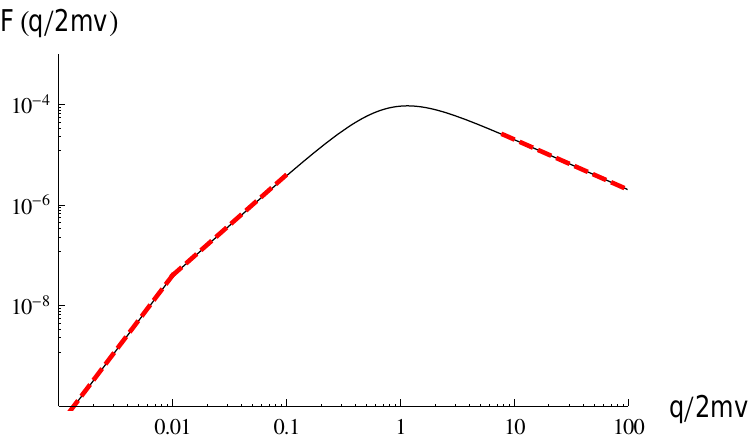}}

\subfloat[$\kf/2mv=20$]{  \includegraphics[clip,width=0.9\columnwidth]{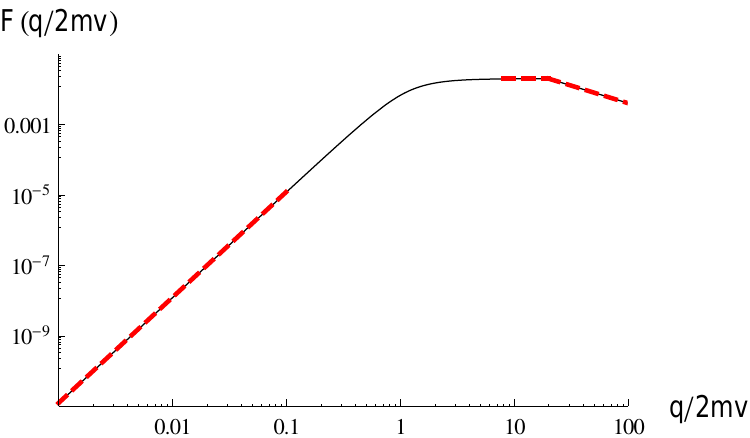}}
\caption{Plot of the crossover function  $F(q/2mv)$ for $G/g=1$, $M/m=100$ for two different values of $\kf/2mv$. The dashed lines are the decay rates (\ref{eq:dr1}) and (\ref{eq:dr3}) in units of ${M G^2}/{2\pi \hbar^3}$.}
\label{fig:dr}
\end{figure}

In Fig.~\ref{fig:dr}, we show the decay rate as a function of $q$ using Eq.~(\ref{eq:drcross}), which is compared with previously obtained results for $q\ll q_0$ and $q\gg q_0$. The decay rate is a nonmonotonic function of the quasiparticle momentum $q$, which reaches the maximum for momenta of the order of $q_0\sim mv$. At the momentum $q=k_F$, the decay rate is a smooth, but nonanalytic function. 

\section{Discussion and conclusion}
\label{sec:stability}

A one-dimensional Bose-Fermi mixture can exhibit different phases at zero temperature \cite{hu2016strongly,mathey_phase_2007, rizzi_pairing_2008, cazalilla_instabilities_2003,das_bose-fermi_2003}. In the case of weak interaction, several approaches \cite{hu2016strongly, rizzi_pairing_2008,cazalilla_instabilities_2003, das_bose-fermi_2003} have shown that a uniform Bose-Fermi mixture is stable against the creation of density fluctuations if the Fermi momentum is not particularly small. This occurs at 
\begin{align}\label{instabilities}
k_F>\frac{MG^2}{\pi\hbar g}.
\end{align} 
Therefore, if the condition (\ref{instabilities}) is satisfied, our system is uniform and the study of the quasiparticle decay is justified. Throughout this paper, we thus assumed that the condition (\ref{instabilities}) is valid. On the other hand, we obtained that our system is well described by the microscopic theory as long as the Fermi momentum is not too high [see Eq.~(\ref{eq:lowkf})], which can be approximately expressed as $k_F<Mv$ at low momenta. The above two conditions on $k_F$ give $G<g\sqrt{K}$, which is always satisfied if one considers a weakly-interacting Bose-Fermi mixture (see the discussion in Appendix \ref{app:eqcorr}). 

We have found that the weakly interacting Bose-Fermi mixture has completely different scaling of the decay rate as a function of the momentum in the case 
\begin{align}\label{qqqq}
\frac{m}{M}=\frac{g}{G}\neq 1.
\end{align} 
This is revealed in the matrix element (\ref{eq:Aexp}), which nullifies under the condition (\ref{qqqq}). However, in this case, one does not expect to have the integrable model. Once we evaluated the matrix element (\ref{eq:Afinal}) under the condition (\ref{qqqq}), we indeed found a nonzero result [see Eq.~(\ref{eq:AexpRr})]. The corresponding decay rate is given by Eq.~(\ref{eq:dr2}), which is the dominant contribution in the region defined by Eq.~(\ref{condmM}).

In this paper, we studied the zero-temperature decay rate of excitations in a one-dimensional weakly interacting Bose-Fermi mixture. The quasiparticles on the Bogoliubov branch of excitations  decay due to the interaction with fermions in the filled Fermi sea. Starting from the hydrodynamic approach, we built a microscopic theory to describe the excitations of the system and their residual interaction. We have shown that the main process of the decay is the backscattering of the Bogoliubov quasiparticle off the Fermi sea producing a particle-hole pair. We calculated the scattering matrix element using the perturbation theory and then, applying the Fermi golden rule, we evaluated the decay rate. The decay rate has different behavior for the momentum of the quasiparticle below or above the Fermi momentum. The reason is the change of  the available phase space for the backscattered Bogoliubov quasiparticle; see Eqs.~(\ref{eq:qMm1}), (\ref{eq:qMm2}), and (\ref{eq:decayratefinal}). This can also be seen in terms of the available phase space for the hole. For the quasiparticle momentum below the Fermi momentum, $q<\kf$, only some part of the states below the Fermi sea can be excited, while in the opposite case, $q\geq\kf$, all the fermionic states below the Fermi sea are excited (this can be easily seen from the consideration done in Appendix \ref{app:tau}). The abrupt change of the Fermi distribution function at $\kf$ is directly seen in the decay rate, which at the momentum $q=\kf$ shows the nonanalytic behavior.

We found that for the low-momentum Bogoliubov quasiparticles, the decay rate scales as $q^3$ at $q<\kf$, while it scales as $q^2$ for $q>\kf$; see \eq{eq:dr1}. We found a dramatic change of the decay rate when the ratio of the masses of fermions and bosons is equal to the ratio of the boson-fermion and boson-boson interaction strengths. In this case, the decay rate scales as $q^7$ for $q<\kf$ and as $q^6$ at higher momenta $q>\kf$; see \eq{eq:dr2}. For quasiparticles at momenta above $mv$, where the Bogoliubov spectrum practically becomes quadratic, we found a momentum-independent decay rate for $q<\kf$, while the rate scales as $1/q$ for $q>\kf$; see \eq{eq:dr4}. We provided the analytic expression for the crossover function describing the decay rate between the mentioned regimes of slow and fast quasiparticles, given by Eq.~(\ref{eq:drcross}). In the integrable case, the decay rate vanishes. Our results could be experimentally detected by measuring the broadening of the dynamic structure factor in one-dimensional Bose-Fermi mixtures.

\appendix
\section{Calculation of the scattering matrix element}
\label{app:amp}

In this appendix, we explain some of the technical details of the calculation leading to \eq{eq:Afinal}. In order to calculate the scattering matrix element $\mathcal A_{i}^{f}$ from an initial state $|i\ra$ to a  final state $|f\ra$ at second-order perturbation theory in $V$, for the standard expression 
\begin{align}
\mathcal A_{i}^{f}=&\sum_m \dfrac{\la f|V|m\ra \la m|V|i \ra}{E_i-E_m}, \label{eq:timeint0}
\end{align}
we use another form,
\begin{align}
\mathcal{A}_{i}^{f}=\lim_{\delta\to0^+}\int_0^\infty{\dfrac{dt}{i\hbar}}\la f|V(0)V(-t)|i\ra e^{-\delta t/\hbar}.\label{eq:timeint}
\end{align}
Here, $E_i=\la i|H_{B0}+H_F|i\ra$ is the energy of the initial state and $E_m=\la m|H_{B0}+H_F|m\ra$ is the energy of the intermediate state. Instead of performing the summation over intermediate states in \eq{eq:timeint0}, we evaluate the time-dependent matrix element $\la f|V(0)V(-t)|i\ra $ and then integrate over time as in \eq{eq:timeint} to obtain the final result for the matrix element. The time evolution of the operators in \eq{eq:timeint} is
	\begin{gather}
		V(t)=e^{it (H_{B0}+H_F)/\hbar}V e^{-it  (H_{B0}+H_F)/\hbar},\\
		b_p(t)=e^{-it\vep_p/\hbar}b_p,\\
		a_p(t)=e^{-itE_p/\hbar}a_p.
	\end{gather}
in the interaction representation.

In the case of  the scattering matrix element (\ref{eq:Ao2}) that we want to evaluate in the main text, the initial and final states are
	\begin{gather}
	|i\ra= a_{k}^\da b_{q}^\da \vak,\\
	|f\ra=a_{k'}^\da b_{q'}^\da \vak,
	\end{gather}
where $\vak$ denotes the vacuum. We now want to calculate $\la f|V(0)V(-t)|i\ra$ and then integrate over time to get the scattering matrix element (\ref{eq:Afinal}). We recall $V=V_1+V_3$. Since  the  number of bosons and fermions in the initial and final state are  the same, respectively, only terms which conserve the number of each particle are kept in the product $V(0)V(-t)$.  These terms are
	\begin{widetext}
	\begin{multline}
		\la f|V(0)V(-t)|i\ra =\displaystyle \sum_{\substack{q_1,q_2,q_3 \\ q_1',q_2',q_3'}} \delta_{q_1'+q_2'+q_3',0}\delta_{q_1+q_2+q_3,0} \left\{\dfrac{3\,|q_1'q_2'q_3'|}{\sqrt{\vep_{q_1'}\vep_{q_2'}\vep_{q_3'}}} \Gamma(q_3) f_-(q_1',q_2',q_3') \vab  a_{k'}  a^\da_{-q_2}a_{q_1} a_{k}^\da\vak   \right.\\
		\times  \left[ \vab b_{q'}b^\da_{-q_1'}b_{q_2'}b_{q_3'}b^\da_{-q_3}b^\da_q \vak e^{-it(E_{q_2}-E_{q_1}+\vep_{q_3})/\hbar}  +\vab b_{q'} b_{q_3}b^\da_{q_3'}b^\da_{q_2'}b_{-q_1'}b^\da_q \vak   e^{-it(\vep_{q_3'}+\vep_{q_2'}-\vep_{q_1'})/\hbar}    \right]\\
\left.		+\Gamma(q_3)\Gamma(q_3') \vab a_{k'} a^\da_{-q_2'}a_{q_1'}  a^\da_{-q_2}a_{q_1} a_{k}^\da\vak \left[ \vab b_{q'}  b^\da_{-q_3'}b_{q_3}  b_q^\da  \vak e^{it\vep_{q_3}/\hbar} + \vab b_{q'}  b_{q_3'}b^\da_{-q_3}  b^\da_q  \vak e^{-it\vep_{q_3}/\hbar}  \right]e^{-it(E_{q_2}-E_{q_1})/\hbar} \vphantom{\dfrac{q_1'}{\sqrt{\vep_{q_1'}}}}\right\}.
	\label{eq:Aint}
	\end{multline}
	\end{widetext}
We now perform  Wick contractions in order to evaluate the different matrix elements. We take care not to  contract $b_q^\da$ with $b_{q'}$ and $a_k^\da$ with $a_{k'}$ since it would imply that the initial boson or the initial fermion did not interact during the process, which is not the process we are considering.  The evaluation of the different matrix elements in \eq{eq:Aint} leads to
	\begin{gather}
		\vab  a_{k'}  a^\da_{-q_2}a_{q_1} a_{k}^\da\vak= \delta_{k,q_1}\delta_{k',-q_2}, \label{eq:wc1}\\
		\vab b_{q'}  b^\da_{-q_3'}b_{q_3}  b_q^\da  \vak= \delta_{q,q_3}\delta_{q',-q_3'},\\
		\vab b_{q'}  b_{q_3'}b^\da_{-q_3}  b^\da_q  \vak= \delta_{q,q_3'}\delta_{q',-q_3},
	\end{gather}
and 
	\begin{equation}
		\vab a_{k'} a^\da_{-q_2'}a_{q_1'}  a^\da_{-q_2}a_{q_1} a_{k}^\da\vak=\delta_{k ,q_1 } \delta_{-q_2 ,q_1' } \delta_{-q_2' , k'}.\label{eq:wc4}
	\end{equation}
For the last two bosonic matrix elements, we use the symmetry of the function $f_-(q_1,q_2,q_3)$ under the permutation $q_2\leftrightarrow q_3$, which implies
	\begin{gather}
		\vab b_{q'}b^\da_{-q_1'}b_{q_2'}b_{q_3'}b^\da_{-q_3}b^\da_q \vak= 2\delta_{q' ,-q_1' }\delta_{q_2' ,q }\delta_{q_3' ,-q_3 },\label{eq:wc5}\\
		\vab b_{q'} b_{q_3}b^\da_{q_3'}b^\da_{q_2'}b_{-q_1'}b^\da_q \vak= 2\delta_{q' ,q_2' }\delta_{q_3' ,q_3 }\delta_{-q_1' ,q }\label{eq:wc6}.
	\end{gather}	
We did not contract $b_{-q_1'}^\da$ with $b_{q_2'}$ or $b_{q_3'}$ in \eq{eq:wc5} because the remaining momentum would cancel due to the conservation of momentum [the same reasoning applies to \eq{eq:wc6}]. We insert Eqs.~(\ref{eq:wc1})--(\ref{eq:wc6}) back into \eq{eq:Aint},  then integrate over time and obtain the scattering matrix element (\ref{eq:Afinal}).

%%%%%%%%%%%%%%%%%%%%%%%%%%%%

%%%%%%%%%%%%%%%%%%%%%%%%
\section{The phase space for $q'$}
\label{q'}
%%%%%%%%%%%%%%%%%%

\begin{figure}
{\subfloat[$\kf<q$]{  \includegraphics[width=0.45\columnwidth]{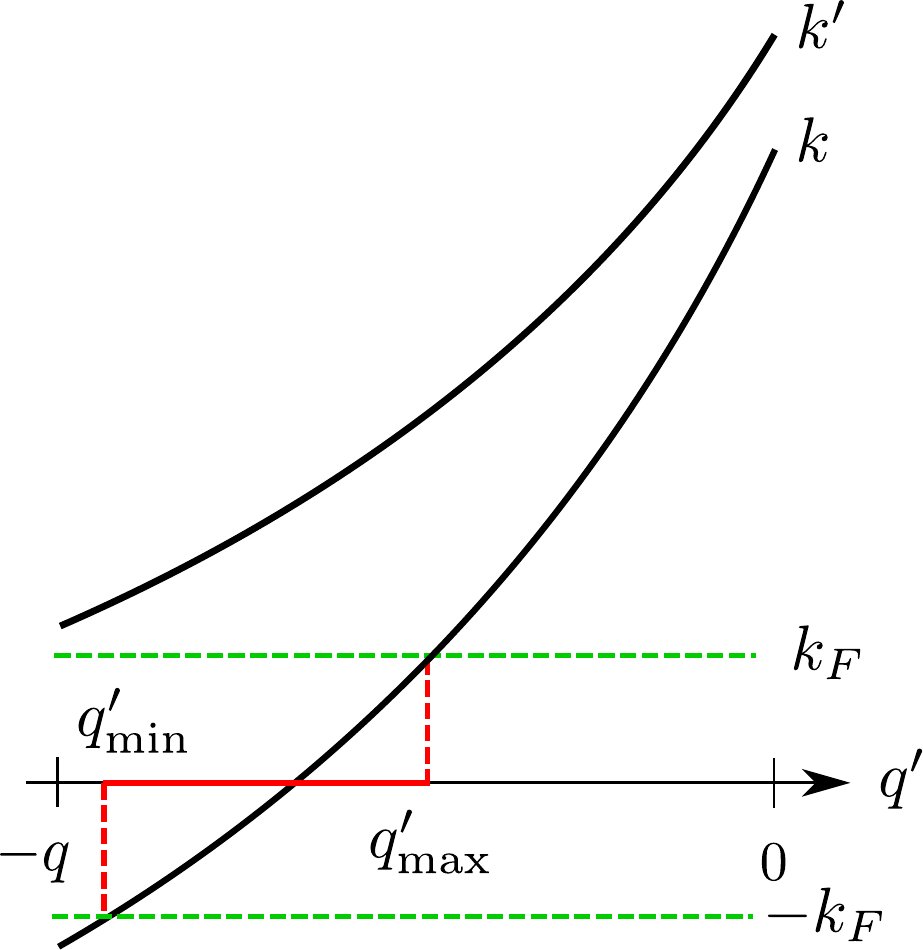}}\label{sfig:k12b}}
\quad
{\subfloat[$\kf>q$]{  \includegraphics[width=0.45\columnwidth]{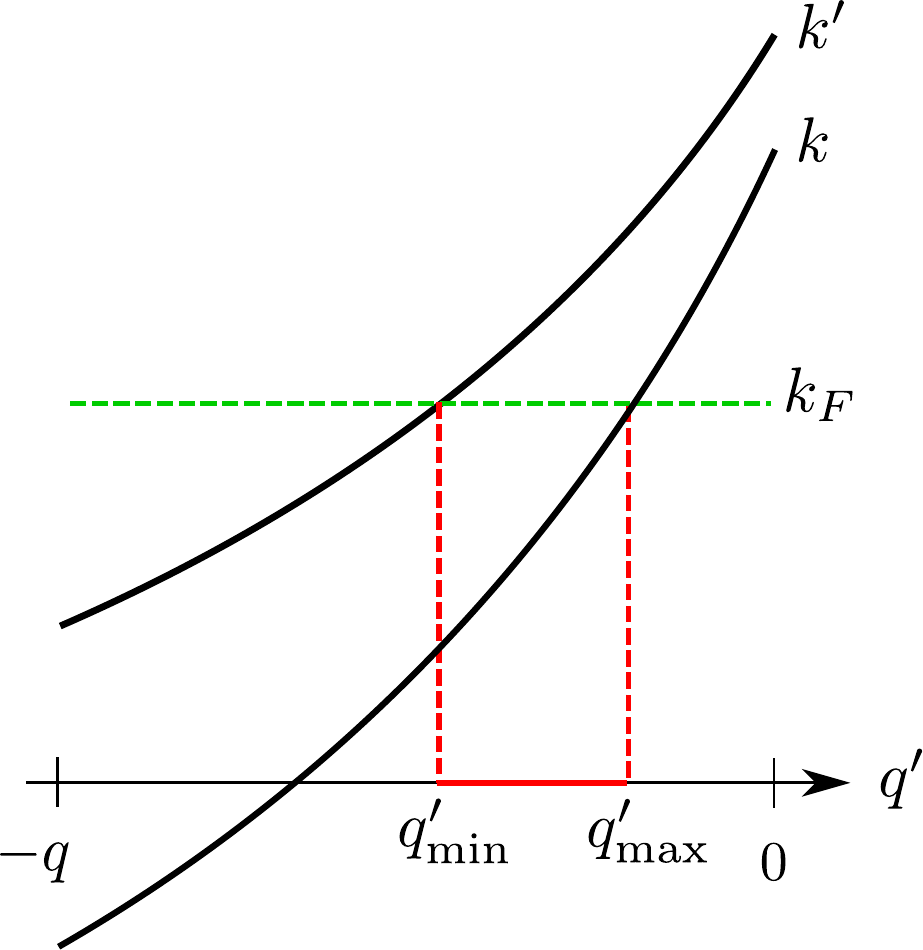}}\label{sfig:k12}}
\caption{Plots of $k'$ and $k$ in the case where $0<q<M v$.  The green dashed line represents the Fermi momentum $\kf$.  (a)  $\kf<q$, (b) $\kf>q$. The red dashed lines determine $q'_\text{min}$ and $q'_\text{max}$. }
\label{fig:k12}
\end{figure}

Using the energy and momentum conservation, we can express $k$ and $k'$ as functions of $q$ and $q'$; see Eqs.~({\ref{k}) and (\ref{k'}}). Here we provide the detailed analysis of the phase space available for $q'$, such that $|k(q,q')|<\kf$ and $|k'(q,q')|>\kf$. We assume $M v>q>0$.  We can show that $k'(q,q')>Mv$ for $0<q'<q$. Since we consider slow fermions  ($|k'|,|k|<Mv$), we are only interested in the region $-q<q'<0$.
The function $k'$  is an increasing function of  $q'$. In the following, we shorten the notation  and keep the explicit dependence only on $q'$, i.e., we use $k(q')$ and $k'(q')$. The minimum of $k'$ is given by  $k'(-q)=q$ and its  maximum by $k'(0)=M\vep_q/q+q/2>Mv$.
We can show that $k$ is an increasing function of $q'$ with a minimum  $k(-q)=-q$ and a maximum $k(0)=M\vep_q/q-q/2>0$. 
One can define two regions, $\kf<q$ and $\kf>q$. In the region $\kf<q$, the condition $k'>\kf$ is automatically satisfied, we only need to satisfy $-\kf<k<\kf$. It implies that the minimal and maximal values of $q'$ are  given by the intersection of $k$ with $\mp \kf$, as given by Eq.~(\ref{eq:qMm1}). In the region $\kf>q$, the condition $-\kf<k$ is automatically satisfied; we only need to satisfy $k'>\kf$ and $k<\kf$. The minimal and maximal values of $q'$ are determined by the intersection of $k'$ and $k$ with $\kf$, as given by Eq.~(\ref{eq:qMm2}). See Fig.~\ref{fig:k12} for a graphic representation of the determination of $q'_\text{min/max}$.

We now discuss the constraint on $\kf$. Since  $k'(q'=0)$ is bigger than $Mv$ and the minimum of $k'$ is less than $Mv$, i.e., $q<Mv$, the equation $k'(q')=Mv$ admits the unique solution  $q'=q'_*$. Since $k'$ is an increasing function of $q'$, $Mv>k'$ implies that the allowed values for $q'$ should be below $q'_*$, and therefore $q'_*>q'_\text{max}$. $k$ is also an increasing function of $q'$ and therefore $k(q'_*)>k(q'_\text{max})=\kf$ [where the last equality comes from Eqs.~(\ref{eq:qMm1}) and (\ref{eq:qMm2}), while it leads to Eq.~(\ref{eq:lowkf})].  Notice that $k(q'_*)$ should be positive, which may impose a constraint on $q$. At large momenta, we find $q<Mv/2$; see Sec.~\ref{CCCC}.

%%%%%%%%%%%%%%%%%%%%%%%%%%%%

\section{An alternative way to obtain decay rate}
\label{app:tau}

Here we give a somewhat different way than the one presented in Sec.~\ref{sec:dr} to obtain the decay rate of a Bogoliubov quasiparticle. We start from the expression for the decay rate (\ref{eq:decayrate})  given by the Fermi golden rule where  we replace the matrix element $\mathcal A_{q,k}^{q',k'}$ by its expression in  \eq{eq:Akro}.
We then  sum over $q'$ to get rid of the Kronecker $\delta$. We obtain
	\be
		\dfrac{1}{\tau}=\frac{2\pi}{\hbar}\sum_{\substack{|k|<\kf\\k'>\kf}}|\Lambda(q,q+k-k')|^2\delta(\vep_q+E_k-\vep_{q+k-k'}-E_{k'}).
	\ee
Note that from conservation of momentum and energy, if $q>0$ then $k'>0$. In order to integrate over $k'$, we now deal with the conservation of energy which is ensured by the Dirac $\delta$,
	\be
		\delta(\vep_q+E_k-\vep_{q+k-k'}-E_{k'})=\left|\partial_{k'}{\vep}_{q+k-{k'}}+\dfrac{{k'}}{M} \right|^{-1}\!\!\delta(k'-{k_0'}).
	\ee
Here $\partial_{k}$ denotes the partial derivative with respect to $k$ and  ${k_0'}$ stands for a function of $q$ and $k$, i.e.,~${k_0'}(k,q)$, that is  defined by the following equation:
	\be
		\vep_q+E_k-\vep_{q+k-{k_0'}}-E_{{k_0'}}=0.
	\ee
The summations are then replaced by integrations over momenta using $\sum_q\rightarrow L/2\pi\hbar \int dq$. After the integration over $k'$, the decay rate takes the following form:
	\be
		\dfrac{1}{\tau}=\dfrac{M L^2}{2\pi\hbar^3}\int_\omega^\kf dk\,\dfrac{|\Lambda(q,q+k-k_0')|^2}{ \left|M\partial_{k_0'}\vep_{q+k-{k_0'}} +{k_0'}\right|}.
	\label{eq:drApp}
	\ee
The lower bound $\omega$ is needed since not all $|k|<\kf$ satisfy the condition $k_0'(k,q)>\kf$. It is given by
	\be
		\omega=
		\begin{cases}
			-\kf&\text{if }q>\kf,\\
			k_0&\text{if }q<\kf,
		\end{cases}
	\ee
where $k_0$ depends on $q$ and is determined by the equation $\kf=k_0'({k_0},q)$. For $q=0$ one finds $k_0=\kf$ and the decay rate vanishes. Also, note that for  $q=\kf$, one has $k_0=-\kf$. Thus the lower bound of integration $\omega$ is the continuous function of $q$, but not a smooth one.

%%%%%%%%%%%%%%%%%%%%%%%%%%%%%%%%%
\subsection{Small momenta}

Now we evaluate the decay rate for $q\ll q_0$.  We can linearize the Bogoliubov dispersion, $\vep_q\simeq vq$ (we consider $q>0$). Since quasiparticles can only be backscattered, we have $\vep_{q'}\simeq- v q'$. The conservation of momentum and energy leads to
\begin{align}
		k_0'&=k+\dfrac{2q Mv}{k+Mv}+\mathcal{O}\left(\dfrac{q^2}{q_0^2} \right), \label{eq:k0psmallq} \\
		k_0&=\kf-\dfrac{2q Mv}{\kf+Mv}+\mathcal{O}\left(\dfrac{q^2}{q_0^2} \right).\label{eq:k0psmallq2}
	\end{align}
In the linear part of the quasiparticle spectrum, the denominator in \eq{eq:drApp} becomes simply $k_0'+Mv$ [where $k_0'$ is given by \eq{eq:k0psmallq}]. We use the scattering matrix element (\ref{eq:Aexp}) in the expression of the decay rate (\ref{eq:drApp}) and obtain, at   leading order in $q/q_0$, the result from \eq{eq:dr1}.
The assumption  on $\kf$   is the same as in the main text.

%%%%%%%%%%%%%%%%%%%%%%%%%%%%%%%%%%
\subsection{Small momenta when $Gm=gM$}

Here, $k_0',k_0$ are the same as in Eqs.~(\ref{eq:k0psmallq}) and (\ref{eq:k0psmallq2}); one only needs to replace the matrix element in  \eq{eq:drApp} by \eq{eq:AexpRr} leading to the decay rate (\ref{eq:dr2}). The assumption  on $\kf$   is the same as in the main text.

%%%%%%%%%%%%%%%%%%

\subsection{Large momenta}

Let us  consider the decay of a quasiparticle with high momentum, $q\gg q_0$. There   we can approximate the Bogoliubov spectrum by the quadratic dispersion, $\vep_q\simeq q^2/2m$. We obtain

	\begin{gather}
		k_0'=\dfrac{k(M-m)+2Mq}{M+m}+\mathcal{O}\left(\frac{q_0^3}{q^3}\right), \label{eq:k0pQ}\\
		k_0=\dfrac{\kf(M+m)-2Mq}{M-m}+\mathcal{O}\left(\frac{q_0^3}{q^3}\right).\label{eq:k0Q}
	\end{gather}
Being in the quadratic part of the excitation spectrum, the denominator in \eq{eq:drApp} becomes $(q M/m-k)[1+\mathcal O(q_0^4/q^4)]$. Then, using the scattering matrix element (\ref{eq:AexpQ}) in the decay rate (\ref{eq:drApp}), we obtain, at leading order,
the same expression as in \eq{eq:dr3}.
The assumptions  on $\kf$ and $M/m$ are the same as in the main text, which allows for the expansion of the decay rate as in \eq{eq:dr4}.
 
%%%%%%%%%%%%%%%%%%%%%%%%
\section{Correction to the energy of fermions due to interaction with bosons}
\label{app:Ekcorr}

In this appendix,  we calculate the leading correction to the  spectrum of   fermions, $\delta E_k$, induced by the interaction with bosons using  perturbation theory. We show that $\delta E_k$ has a divergent behavior as $k$ approaches $Mv$. This divergence  indicates the breakdown of perturbation theory and therefore allows us to describe only slow fermions.

The energy of a fermionic excitation interacting with bosons is given by
	\be
		E_k^\text{Tot}=E_k+Gn_0+\delta E_k+\ldots
	\ee
where $\delta E_k$ is the leading correction to the fermionic spectrum. The first-order perturbation in $V=V_1+V_3$ vanishes (i.e., $\la k|V |k\ra=0$) since  the initial and final states do not contain any bosons and $V$ does not conserve the number of bosons. Therefore, we look for the leading correction in second-order perturbation theory: 
	\be
		\delta E_k=\sum_m \dfrac{\la k|V |m \ra\la m|V|k\ra}{E_{|k\ra}-E_m},
		\label{eq:dEk}
	\ee
where $E_{|k\ra}=\la k|H_F|k\ra$ and $E_m=\la m|H_{B0}+H_F|m\ra$. We consider an excitation above the Fermi sea ($k>\kf$) since it is relevant to the process studied in the main text. The state $|k\ra$  is defined by
	\be
		|k\ra=a_k^\da\fsk,\quad\fsk=\prod_{|q|\leq\kf}a_q^\da\vak.
		\label{eq:fs}
	\ee
We did not use the Fermi sea as background up to now in this paper; however, the results we obtained for the different scattering matrix elements are the same up to a global factor $\pm1$ depending on how the initial and final states are defined. 
Since \eq{eq:dEk} has the same structure as   \eq{eq:timeint0} with $|i\ra=|f\ra=|k\ra$, we use the same method as in \eq{eq:timeint} to evaluate $\delta E_k$.  We want to  evaluate $\la k|V(0)V(-t)|k\ra$ and then integrate over time to obtain $\delta E_k$. The only nonzero term arises from $V_1$ perturbation, $\la k|V_1(0)V_1(-t)|k\ra$. It is given by
	\begin{align}
		\la k|&V_1(0)V_1(-t)|k\ra=\displaystyle \sum_{\substack{q_1,q_2,q_3 \\ q_1',q_2',q_3'}} \delta_{q_1'+q_2'+q_3',0}\delta_{q_1+q_2+q_3,0}\notag\\
		&\times\Gamma(q_3)\Gamma(q_3')\vab b_{q_3'}b_{-q_3}^\da\vak e^{-it(E_{q_2}-E_{q_1}+\vep_{q_3})/\hbar}\notag\\
		&\times\fsb a_k a_{-q_2'}^\da a_{q_1'} a_{-q_2}^\da a_{q_1}a_{k}^\da \fsk.
	\end{align}
We now perform Wick contractions. The bosonic matrix element is simply $\vab b_{q_3'}b_{-q_3}^\da\vak=\delta_{q_3',-q_3}$. For the matrix element  involving fermionic operators, we do not contract $a_k$ with $a_k^\da$ since it would imply that the initial fermion did not interaction with bosons. We do not contract $a_{q_1}$ with $a_{-q_2}^\da$ (or $a_{q_1'}$ and $a_{-q_2'}^\da$) because it would imply $q_3=0$ ($q_3'=0$).  The fermionic matrix element  reads
	\begin{align}
		\fsb a_k & a_{-q_2'}^\da a_{q_1'} a_{-q_2}^\da a_{q_1}a_{k}^\da \fsk\notag\\
		=&\,\theta(|q_2|-\kf)\delta_{k,-q_2'}\delta_{q_1',-q_2 }\delta_{q_1 ,k }\notag\\
		&-\theta(\kf-|q_1|)\delta_{k ,-q_2 }\delta_{-q_2' , q_1}\delta_{q_1' ,k }
	\end{align}
where the Heaviside functions $\theta$ account for the Fermi sea at zero temperature. After the integration over time, one obtains
	\begin{align}
		&\delta E_k=\sum_q\frac{ \Gamma(q)^2}{E_k-E_{k-q}-\vep_q}\notag\\
		&-\sum_{q\atop |k-q|<\kf}\!\!\!\left(\dfrac{\Gamma(q)^2 }{E_k-E_{k-q}-\vep_q}-\dfrac{\Gamma(q)^2 }{E_k-E_{k-q}+\vep_q}\right) .
		\label{eq:dEk2}
	\end{align}
We define the first line of the previous equation by $\delta E_k^0=\delta E_k|_{\kf=0}$, while the remaining term we denote by $\delta E_k^F=\delta E_k-\delta E_k^0$.
The expression (\ref{eq:dEk2}) can be evaluated exactly, as we explain now. We start with $\delta E_k^0$. Transforming the sum into an integral $\sum_q\to L/2\pi\hbar\int dq$, and using the change of variable $q\to 2 mv \sinh{x}$, one obtains 
	\begin{equation}
	\delta E_k^0=\dfrac{m v^2}{2K}\dfrac{G^2}{g^2}I,
	\end{equation}
where
	\be
	I=\int_{-\infty}^\infty{dx \left[{\text{sgn}(x)\left(a-b\,\sinh{x} \right)-\cosh{x}}\right]^{-1}},
	\ee
with $a=k/Mv$ and $b=m/M$ for notational convenience. Using the change of variable $x\to \ln t$, one has
	\begin{align}
		I=&-2\int_0^1dt \left[1+b+2 a t+(1-b)t^2 \right]^{-1}\notag\\
		&+2\int_1^\infty dt \left[-1+b+2 a t-(1+b)t^2 \right]^{-1}\notag\\
		=&-\dfrac{2}{\Delta}\dfrac{M}{m}\textrm{arctanh}(\Delta),
	\end{align}
where
	\be\label{Delta}
		\Delta=\sqrt{1-\dfrac{M^2}{m^2}\left[1-\left(\dfrac{k}{Mv} \right)^2\right]}.
	\ee
For the second term, we again use the change of variable $q\to 2 mv \sinh{x}$, and obtain 
	\be
		\delta E_k^F=-\dfrac{m v^2}{2K}\dfrac{G^2}{g^2}I',
	\ee
where
	\begin{align}
		I'=\int_{x_-}^{x_+} dx&\left[ \left( a-b\,\sinh{x}-\cosh{x}\right)^{-1} \right.\notag\\
			&\left. +\left( -a+b\,\sinh{x}-\cosh{x}\right)^{-1} \right].
	\end{align}
The bounds of integration are $x_\pm=\sinh^{-1}[(k\pm\kf)/2mv]>0$ since $k>\kf$. Now we perform the change of variable $x\to\ln t$ and have
	\begin{align}
		I'=&-2\int_{t_-}^{t_+}dt \left[\left(-1+b+2 a t-(1+b)t^2 \right)^{-1}\right.\notag\\
		&\hphantom{-2\int_{t_-}^{t_+}dt }\left.- \left(1+b+2 a t+(1-b)t^2\right)^{-1}\right],
\end{align}	
where $t_\pm=(k\pm\kf)/2mv+\sqrt{1+[(k\pm\kf)/2mv]^2}$. We finally obtain
	\begin{align}	
		I'=-\dfrac{2}{\Delta}\dfrac{M}{m}\textrm{arctanh}\left( \dfrac{4\kf Mv\Delta}{(k^2-\kf^2)\left(\frac{M^2}{m^2}-1\right)+4M^2v^2}\right).
	\end{align}
The leading correction to the  fermionic spectrum  eventually reads
	\begin{align}
	\delta E_k=&-\dfrac{M v^2}{K \Delta}\dfrac{G^2}{g^2}\left[ \textrm{arctanh}(\Delta)  \vphantom{\dfrac{4\kf Mv\Delta}{4M^2v^2}} \right.\notag \\ 
		&\left.-\textrm{arctanh}\left( \dfrac{4\kf Mv\Delta}{(k^2-\kf^2)\left(\frac{M^2}{m^2}-1\right)+4M^2v^2}\right)\right],\label{eq:dekfinal}
	\end{align}
where we recall the definition for $\Delta$ in \eq{Delta}.

One has to notice that   $\delta E_k^0$ diverges as $k$ approaches $Mv$, which signifies a breakdown of the perturbation theory  since  $\delta E_k$ becomes of the same order as $E_k$ at momenta very close to $Mv$. This signals that the fermionic spectrum $E_k=k^2/2M$ is not correct at such high momenta. This allows us   to use microscopic theory only to describe slow fermions of momenta $k<Mv$. The latter condition is actually a simplified version of a more precise condition $k<Mv(1-\alpha)$, where $\alpha>0$ is very small and thus neglected for our purposes.

In the process studied in the main text, a Bogoliubov quasiparticle  is backscattered and excites a particle-hole pair ($k'>\kf$, $|k|<\kf$). In order to avoid the problem of the divergence of $\delta E_k^0$ when $k'\to Mv$, one should consider fermions $k'<Mv$, which constrains the Fermi momentum $\kf$ to small values, as discussed in Appendix~\ref{q'}. The second term, $\delta E_k^F$, is finite if we consider slow fermions.

%%%%%%%%%%%%%%%%%%%%%%%%
\section{Correction to the energy of Bogoliubov quasiparticles due to interaction with the Fermi sea}
\label{app:eqcorr}

In this appendix,  we calculate the leading correction to the spectrum of Bogoliubov quasiparticles, $\delta \vep_q^F$, due to their interaction with the Fermi sea at zero temperature. We also discuss  the leading correction to the sound velocity due to the same interaction, $\delta v^F$.

The  leading  correction to $\varepsilon_q$ is obtained in second order, 
	\be
		\delta\vep_q^F=\sum_m\dfrac{\la q|V_1|m\ra\la m|V_1|q\ra}{E_{|q\ra}-E_m},
		\label{eq:dvepq}
	\ee
where $E_{|q\ra}=\la q|H_{B0}+H_F| q\ra$ and $E_m=\la m|H_{B0}+H_F|m\ra$. The initial and final states are $|q\ra=b_q^\da\vak\fsk$, where $\fsk$ is the Fermi sea defined in \eq{eq:fs}. Equation~(\ref{eq:dvepq}) has the same structure as \eq{eq:timeint0} with $|i\ra=|f\ra=|q\ra$. We thus evaluate
	\begin{align}
		&\la q|V_1(0)V_1(-t)|q\ra=\sum_{\substack{q_1,q_2,q_3 \\ q_1',q_2',q_3'}} \delta_{q_1'+q_2'+q_3',0}\delta_{q_1+q_2+q_3,0}\notag\\
		&\times  \Gamma(q_3')\Gamma(q_3) \fsb  a_{-q_2'}^\da a_{q_1'} a_{-q_2}^\da a_{q_1} \fsk e^{-it(E_{q_2}-E_{q_1})/\hbar}\notag\\
		&\times \left(   \vab b_q  b_{-q_3'}^\da b_{q_3}  b_q^\da \vak e^{it\vep_{q_3}/\hbar} + \vab b_q b_{q_3'} b_{-q_3}^\da    b_q^\da \vak e^{-it\vep_{q_3}/\hbar}\right). \label{eq:eqtmp}
	\end{align}
We now perform Wick contractions. For the fermionic matrix element, we have
	\begin{multline}
		\fsb  a_{-q_2'}^\da a_{q_1'} a_{-q_2}^\da a_{q_1} \fsk =\\
			 \theta(\kf-|q_1|)\theta(|q_2|-\kf)\delta_{q_1,-q_2'}\delta_{q_1',-q_2}, \label{eq:1}
	\end{multline}
where we did not contract $a_{-q_2}^\da$ with $a_{q_1}$ (or $a_{-q_2'}^\da$ with $a_{q_1'}$) since it would imply $q_3=0$ ($q_3'=0$) because of the conservation of momentum. The Heaviside functions account for the Fermi sea.
The bosonic matrix elements are given by
	\begin{gather}
		\vab b_q  b_{-q_3'}^\da b_{q_3}  b_q^\da \vak=\delta_{q,q_3}\delta_{q,-q_3'},\label{eq:2}\\
		\vab b_q b_{q_3'} b_{-q_3}^\da    b_q^\da \vak=\delta_{q,-q_3}\delta_{q,q_3'},\label{eq:3}
	\end{gather}
where we did not contract $b_q^\da$ with $b_q$.
Inserting Eqs.~(\ref{eq:1})--(\ref{eq:3}) into \eq{eq:eqtmp} and integrating over time, one obtains 
	\begin{align}
		\delta\vep_q^F=&-\Gamma(q)^2\sum_p\theta(\kf-|p|)\theta(|q+p|-\kf)\notag\\
		&\times\left(\dfrac{1}{E_{q+p}-E_p-\vep_q}+ \dfrac{1}{E_{q+p}-E_p+\vep_q}\right).
	\end{align}
After integration over $p$, the correction to the energy of Bogoliubov quasiparticles due to the interaction with the Fermi sea can be written in the following form:
	\begin{align}
		\delta\vep_q^F={}&\frac{\varepsilon_q}{2K}\frac{ MG^2}{mg^2}\frac{mv}{|q|}\frac{1}{1+\frac{q^2}{4m^2v^2}}\notag\\
		&\times\textrm{arctanh}\left(\dfrac{\kf |q|}{M^2v^2-\kf^2-\frac{q^2}{4}\left(1-\frac{M^2}{m^2}\right)}\right). \label{eq:eqcorr}
	\end{align}
The correction $\delta\vep_q^F$ in Eq.~(\ref{eq:eqcorr}) is small unless the momentum of the Bogoliubov quasiparticle is high. We find that $\delta\vep_q^F$ diverges at  
\be
q=2\dfrac{\kf-(M/m) \sqrt{\kf^2+m^2v^2[1-(M/m)^2]}}{(M/m)^2-1},
	\label{eq:deqdiv}
\ee
where we assumed $q>0$. We note that at small Fermi momenta, $k_F<Mv\sqrt{1-m^2/M^2}$, the correction (\ref{eq:eqcorr}) does not diverge at any $q$, and thus it is small at weak interaction.

In the context of the process   studied in the main text, this correction is always finite since we consider slow fermions. Indeed one can see that \eq{eq:eqcorr} diverges for $\kf=M\vep_q/q- q/2$ which   corresponds to   $\kf=k(q'=0)$. However the intersection of  $\kf$ with $k$ determines $q_\text{max}'$ (see Appendix~\ref{q'}) and $\kf=k(q'=0)$ would imply $q'_\text{max}=0$. Since we consider slow fermions ($k',|k|<Mv$), $q'_\text{max}<0$ during the process, therefore $\delta\vep_q^F$ is finite.

From  \eq{eq:eqcorr}, one can obtain the leading correction to the sound velocity of bosons due to their interaction with the Fermi sea. The sound velocity $v$ is given by the slope of the Bogoliubov spectrum at $q\to 0$. The correction to $v$ from the interaction with fermions is given by
	\be
		\delta v^F=\partial_q\delta\vep_q^F\bigg{|}_{q\to 0}=\dfrac{v_F}{2K}\dfrac{G^2}{g^2}\frac{1}{1-\left( \dfrac{v_F}{v}\right)^2}, \label{eq:vcorr}
	\ee
where $v_F=k_F/M$. We note that $\delta v^F\ll v$  for $v_F<v$ and at weak interaction. The latter occurs at Bose-Fermi coupling $G\ll g\sqrt{K}$. We also require the condition $K\gg 1$ to have a weakly interacting bosonic subsystem, which is well described by the Bogoliubov excitation spectrum. 

%\nocite{*}
%\bibliography{biblio}

\begin{thebibliography}{47}%
	\makeatletter
	\providecommand \@ifxundefined [1]{%
		\@ifx{#1\undefined}
	}%
	\providecommand \@ifnum [1]{%
		\ifnum #1\expandafter \@firstoftwo
		\else \expandafter \@secondoftwo
		\fi
	}%
	\providecommand \@ifx [1]{%
		\ifx #1\expandafter \@firstoftwo
		\else \expandafter \@secondoftwo
		\fi
	}%
	\providecommand \natexlab [1]{#1}%
	\providecommand \enquote  [1]{``#1''}%
	\providecommand \bibnamefont  [1]{#1}%
	\providecommand \bibfnamefont [1]{#1}%
	\providecommand \citenamefont [1]{#1}%
	\providecommand \href@noop [0]{\@secondoftwo}%
	\providecommand \href [0]{\begingroup \@sanitize@url \@href}%
	\providecommand \@href[1]{\@@startlink{#1}\@@href}%
	\providecommand \@@href[1]{\endgroup#1\@@endlink}%
	\providecommand \@sanitize@url [0]{\catcode `\\12\catcode `\$12\catcode
		`\&12\catcode `\#12\catcode `\^12\catcode `\_12\catcode `\%12\relax}%
	\providecommand \@@startlink[1]{}%
	\providecommand \@@endlink[0]{}%
	\providecommand \url  [0]{\begingroup\@sanitize@url \@url }%
	\providecommand \@url [1]{\endgroup\@href {#1}{\urlprefix }}%
	\providecommand \urlprefix  [0]{URL }%
	\providecommand \Eprint [0]{\href }%
	\providecommand \doibase [0]{http://dx.doi.org/}%
	\providecommand \selectlanguage [0]{\@gobble}%
	\providecommand \bibinfo  [0]{\@secondoftwo}%
	\providecommand \bibfield  [0]{\@secondoftwo}%
	\providecommand \translation [1]{[#1]}%
	\providecommand \BibitemOpen [0]{}%
	\providecommand \bibitemStop [0]{}%
	\providecommand \bibitemNoStop [0]{.\EOS\space}%
	\providecommand \EOS [0]{\spacefactor3000\relax}%
	\providecommand \BibitemShut  [1]{\csname bibitem#1\endcsname}%
	\let\auto@bib@innerbib\@empty
	%</preamble>
	\bibitem [{\citenamefont {Bloch}\ \emph {et~al.}(2008)\citenamefont {Bloch},
		\citenamefont {Dalibard},\ and\ \citenamefont
		{Zwerger}}]{bloch2008many-body}%
	\BibitemOpen
	\bibfield  {author} {\bibinfo {author} {\bibfnamefont {I.}~\bibnamefont
			{Bloch}}, \bibinfo {author} {\bibfnamefont {J.}~\bibnamefont {Dalibard}}, \
		and\ \bibinfo {author} {\bibfnamefont {W.}~\bibnamefont {Zwerger}},\ }\href
	{\doibase 10.1103/RevModPhys.80.885} {\bibfield  {journal} {\bibinfo
			{journal} {Reviews of Modern Physics}\ }\textbf {\bibinfo {volume} {80}},\
		\bibinfo {pages} {885} (\bibinfo {year} {2008})}\BibitemShut {NoStop}%
	\bibitem [{\citenamefont {Schreck}\ \emph {et~al.}(2001)\citenamefont
		{Schreck}, \citenamefont {Khaykovich}, \citenamefont {Corwin}, \citenamefont
		{Ferrari}, \citenamefont {Bourdel}, \citenamefont {Cubizolles},\ and\
		\citenamefont {Salomon}}]{schreck2001quasipure}%
	\BibitemOpen
	\bibfield  {author} {\bibinfo {author} {\bibfnamefont {F.}~\bibnamefont
			{Schreck}}, \bibinfo {author} {\bibfnamefont {L.}~\bibnamefont {Khaykovich}},
		\bibinfo {author} {\bibfnamefont {K.~L.}\ \bibnamefont {Corwin}}, \bibinfo
		{author} {\bibfnamefont {G.}~\bibnamefont {Ferrari}}, \bibinfo {author}
		{\bibfnamefont {T.}~\bibnamefont {Bourdel}}, \bibinfo {author} {\bibfnamefont
			{J.}~\bibnamefont {Cubizolles}}, \ and\ \bibinfo {author} {\bibfnamefont
			{C.}~\bibnamefont {Salomon}},\ }\href {\doibase
		10.1103/PhysRevLett.87.080403} {\bibfield  {journal} {\bibinfo  {journal}
			{Physical Review Letters}\ }\textbf {\bibinfo {volume} {87}},\ \bibinfo
		{pages} {080403} (\bibinfo {year} {2001})}\BibitemShut {NoStop}%
	\bibitem [{\citenamefont {Hadzibabic}\ \emph {et~al.}(2002)\citenamefont
		{Hadzibabic}, \citenamefont {Stan}, \citenamefont {Dieckmann}, \citenamefont
		{Gupta}, \citenamefont {Zwierlein}, \citenamefont {G{\"o}rlitz},\ and\
		\citenamefont {Ketterle}}]{hadzibabic2002two-species}%
	\BibitemOpen
	\bibfield  {author} {\bibinfo {author} {\bibfnamefont {Z.}~\bibnamefont
			{Hadzibabic}}, \bibinfo {author} {\bibfnamefont {C.~A.}\ \bibnamefont
			{Stan}}, \bibinfo {author} {\bibfnamefont {K.}~\bibnamefont {Dieckmann}},
		\bibinfo {author} {\bibfnamefont {S.}~\bibnamefont {Gupta}}, \bibinfo
		{author} {\bibfnamefont {M.~W.}\ \bibnamefont {Zwierlein}}, \bibinfo {author}
		{\bibfnamefont {A.}~\bibnamefont {G{\"o}rlitz}}, \ and\ \bibinfo {author}
		{\bibfnamefont {W.}~\bibnamefont {Ketterle}},\ }\href {\doibase
		10.1103/PhysRevLett.88.160401} {\bibfield  {journal} {\bibinfo  {journal}
			{Physical Review Letters}\ }\textbf {\bibinfo {volume} {88}},\ \bibinfo
		{pages} {160401} (\bibinfo {year} {2002})}\BibitemShut {NoStop}%
	\bibitem [{\citenamefont {Modugno}\ \emph {et~al.}(2002)\citenamefont
		{Modugno}, \citenamefont {Modugno}, \citenamefont {Riboli}, \citenamefont
		{Roati},\ and\ \citenamefont {Inguscio}}]{modugno2002two}%
	\BibitemOpen
	\bibfield  {author} {\bibinfo {author} {\bibfnamefont {G.}~\bibnamefont
			{Modugno}}, \bibinfo {author} {\bibfnamefont {M.}~\bibnamefont {Modugno}},
		\bibinfo {author} {\bibfnamefont {F.}~\bibnamefont {Riboli}}, \bibinfo
		{author} {\bibfnamefont {G.}~\bibnamefont {Roati}}, \ and\ \bibinfo {author}
		{\bibfnamefont {M.}~\bibnamefont {Inguscio}},\ }\href {\doibase
		10.1103/PhysRevLett.89.190404} {\bibfield  {journal} {\bibinfo  {journal}
			{Physical Review Letters}\ }\textbf {\bibinfo {volume} {89}},\ \bibinfo
		{pages} {190404} (\bibinfo {year} {2002})}\BibitemShut {NoStop}%
	\bibitem [{\citenamefont {Goldwin}\ \emph {et~al.}(2004)\citenamefont
		{Goldwin}, \citenamefont {Inouye}, \citenamefont {Olsen}, \citenamefont
		{Newman}, \citenamefont {DePaola},\ and\ \citenamefont
		{Jin}}]{goldwin2004measurement}%
	\BibitemOpen
	\bibfield  {author} {\bibinfo {author} {\bibfnamefont {J.}~\bibnamefont
			{Goldwin}}, \bibinfo {author} {\bibfnamefont {S.}~\bibnamefont {Inouye}},
		\bibinfo {author} {\bibfnamefont {M.~L.}\ \bibnamefont {Olsen}}, \bibinfo
		{author} {\bibfnamefont {B.}~\bibnamefont {Newman}}, \bibinfo {author}
		{\bibfnamefont {B.~D.}\ \bibnamefont {DePaola}}, \ and\ \bibinfo {author}
		{\bibfnamefont {D.~S.}\ \bibnamefont {Jin}},\ }\href {\doibase
		10.1103/PhysRevA.70.021601} {\bibfield  {journal} {\bibinfo  {journal}
			{Physical Review A}\ }\textbf {\bibinfo {volume} {70}},\ \bibinfo {pages}
		{021601} (\bibinfo {year} {2004})}\BibitemShut {NoStop}%
	\bibitem [{\citenamefont {Ospelkaus}\ \emph {et~al.}(2006)\citenamefont
		{Ospelkaus}, \citenamefont {Ospelkaus}, \citenamefont {Sengstock},\ and\
		\citenamefont {Bongs}}]{ospelkaus2006interaction-driven}%
	\BibitemOpen
	\bibfield  {author} {\bibinfo {author} {\bibfnamefont {C.}~\bibnamefont
			{Ospelkaus}}, \bibinfo {author} {\bibfnamefont {S.}~\bibnamefont
			{Ospelkaus}}, \bibinfo {author} {\bibfnamefont {K.}~\bibnamefont
			{Sengstock}}, \ and\ \bibinfo {author} {\bibfnamefont {K.}~\bibnamefont
			{Bongs}},\ }\href {\doibase 10.1103/PhysRevLett.96.020401} {\bibfield
		{journal} {\bibinfo  {journal} {Physical Review Letters}\ }\textbf {\bibinfo
			{volume} {96}},\ \bibinfo {pages} {020401} (\bibinfo {year}
		{2006})}\BibitemShut {NoStop}%
	\bibitem [{\citenamefont {Ferrier-Barbut}\ \emph {et~al.}(2014)\citenamefont
		{Ferrier-Barbut}, \citenamefont {Delehaye}, \citenamefont {Laurent},
		\citenamefont {Grier}, \citenamefont {Pierce}, \citenamefont {Rem},
		\citenamefont {Chevy},\ and\ \citenamefont
		{Salomon}}]{ferrier-barbut2014mixture}%
	\BibitemOpen
	\bibfield  {author} {\bibinfo {author} {\bibfnamefont {I.}~\bibnamefont
			{Ferrier-Barbut}}, \bibinfo {author} {\bibfnamefont {M.}~\bibnamefont
			{Delehaye}}, \bibinfo {author} {\bibfnamefont {S.}~\bibnamefont {Laurent}},
		\bibinfo {author} {\bibfnamefont {A.~T.}\ \bibnamefont {Grier}}, \bibinfo
		{author} {\bibfnamefont {M.}~\bibnamefont {Pierce}}, \bibinfo {author}
		{\bibfnamefont {B.~S.}\ \bibnamefont {Rem}}, \bibinfo {author} {\bibfnamefont
			{F.}~\bibnamefont {Chevy}}, \ and\ \bibinfo {author} {\bibfnamefont
			{C.}~\bibnamefont {Salomon}},\ }\href {\doibase 10.1126/science.1255380}
	{\bibfield  {journal} {\bibinfo  {journal} {Science}\ }\textbf {\bibinfo
			{volume} {345}},\ \bibinfo {pages} {1035} (\bibinfo {year}
		{2014})}\BibitemShut {NoStop}%
	\bibitem [{\citenamefont {Lai}\ and\ \citenamefont
		{Yang}(1971)}]{lai1971ground-state}%
	\BibitemOpen
	\bibfield  {author} {\bibinfo {author} {\bibfnamefont {C.~K.}\ \bibnamefont
			{Lai}}\ and\ \bibinfo {author} {\bibfnamefont {C.~N.}\ \bibnamefont {Yang}},\
	}\href {\doibase 10.1103/PhysRevA.3.393} {\bibfield  {journal} {\bibinfo
		{journal} {Physical Review A}\ }\textbf {\bibinfo {volume} {3}},\ \bibinfo
	{pages} {393} (\bibinfo {year} {1971})}\BibitemShut {NoStop}%
\bibitem [{\citenamefont {Das}(2003)}]{das_bose-fermi_2003}%
\BibitemOpen
\bibfield  {author} {\bibinfo {author} {\bibfnamefont {K.~K.}\ \bibnamefont
		{Das}},\ }\href {\doibase 10.1103/PhysRevLett.90.170403} {\bibfield
	{journal} {\bibinfo  {journal} {Physical Review Letters}\ }\textbf {\bibinfo
		{volume} {90}},\ \bibinfo {pages} {170403} (\bibinfo {year}
	{2003})}\BibitemShut {NoStop}%
\bibitem [{\citenamefont {Cazalilla}\ and\ \citenamefont
	{Ho}(2003)}]{cazalilla_instabilities_2003}%
\BibitemOpen
\bibfield  {author} {\bibinfo {author} {\bibfnamefont {M.~A.}\ \bibnamefont
		{Cazalilla}}\ and\ \bibinfo {author} {\bibfnamefont {A.~F.}\ \bibnamefont
		{Ho}},\ }\href {\doibase 10.1103/PhysRevLett.91.150403} {\bibfield  {journal}
	{\bibinfo  {journal} {Physical Review Letters}\ }\textbf {\bibinfo {volume}
		{91}},\ \bibinfo {pages} {150403} (\bibinfo {year} {2003})}\BibitemShut
{NoStop}%
\bibitem [{\citenamefont {Mathey}\ \emph {et~al.}(2004)\citenamefont {Mathey},
	\citenamefont {Wang}, \citenamefont {Hofstetter}, \citenamefont {Lukin},\
	and\ \citenamefont {Demler}}]{mathey2004luttinger}%
\BibitemOpen
\bibfield  {author} {\bibinfo {author} {\bibfnamefont {L.}~\bibnamefont
		{Mathey}}, \bibinfo {author} {\bibfnamefont {D.-W.}\ \bibnamefont {Wang}},
	\bibinfo {author} {\bibfnamefont {W.}~\bibnamefont {Hofstetter}}, \bibinfo
	{author} {\bibfnamefont {M.~D.}\ \bibnamefont {Lukin}}, \ and\ \bibinfo
	{author} {\bibfnamefont {E.}~\bibnamefont {Demler}},\ }\href {\doibase
	10.1103/PhysRevLett.93.120404} {\bibfield  {journal} {\bibinfo  {journal}
		{Physical Review Letters}\ }\textbf {\bibinfo {volume} {93}},\ \bibinfo
	{pages} {120404} (\bibinfo {year} {2004})}\BibitemShut {NoStop}%
\bibitem [{\citenamefont {Takeuchi}\ and\ \citenamefont
	{Mori}(2005)}]{takeuchi_mixing-demixing_2005}%
\BibitemOpen
\bibfield  {author} {\bibinfo {author} {\bibfnamefont {Y.}~\bibnamefont
		{Takeuchi}}\ and\ \bibinfo {author} {\bibfnamefont {H.}~\bibnamefont
		{Mori}},\ }\href {\doibase 10.1103/PhysRevA.72.063617} {\bibfield  {journal}
	{\bibinfo  {journal} {Physical Review A}\ }\textbf {\bibinfo {volume} {72}},\
	\bibinfo {pages} {063617} (\bibinfo {year} {2005})}\BibitemShut {NoStop}%
\bibitem [{\citenamefont {Batchelor}\ \emph {et~al.}(2005)\citenamefont
	{Batchelor}, \citenamefont {Bortz}, \citenamefont {Guan},\ and\ \citenamefont
	{Oelkers}}]{batchelor_exact_2005}%
\BibitemOpen
\bibfield  {author} {\bibinfo {author} {\bibfnamefont {M.~T.}\ \bibnamefont
		{Batchelor}}, \bibinfo {author} {\bibfnamefont {M.}~\bibnamefont {Bortz}},
	\bibinfo {author} {\bibfnamefont {X.~W.}\ \bibnamefont {Guan}}, \ and\
	\bibinfo {author} {\bibfnamefont {N.}~\bibnamefont {Oelkers}},\ }\href
{\doibase 10.1103/PhysRevA.72.061603} {\bibfield  {journal} {\bibinfo
		{journal} {Physical Review A}\ }\textbf {\bibinfo {volume} {72}},\ \bibinfo
	{pages} {061603} (\bibinfo {year} {2005})}\BibitemShut {NoStop}%
\bibitem [{\citenamefont {Frahm}\ and\ \citenamefont
	{Palacios}(2005)}]{frahm2005correlation}%
\BibitemOpen
\bibfield  {author} {\bibinfo {author} {\bibfnamefont {H.}~\bibnamefont
		{Frahm}}\ and\ \bibinfo {author} {\bibfnamefont {G.}~\bibnamefont
		{Palacios}},\ }\href {\doibase 10.1103/PhysRevA.72.061604} {\bibfield
	{journal} {\bibinfo  {journal} {Physical Review A}\ }\textbf {\bibinfo
		{volume} {72}},\ \bibinfo {pages} {061604} (\bibinfo {year}
	{2005})}\BibitemShut {NoStop}%
\bibitem [{\citenamefont {Imambekov}\ and\ \citenamefont
	{Demler}(2006{\natexlab{a}})}]{imambekov2006exactly}%
\BibitemOpen
\bibfield  {author} {\bibinfo {author} {\bibfnamefont {A.}~\bibnamefont
		{Imambekov}}\ and\ \bibinfo {author} {\bibfnamefont {E.}~\bibnamefont
		{Demler}},\ }\href {\doibase 10.1103/PhysRevA.73.021602} {\bibfield
	{journal} {\bibinfo  {journal} {Physical Review A}\ }\textbf {\bibinfo
		{volume} {73}},\ \bibinfo {pages} {021602} (\bibinfo {year}
	{2006}{\natexlab{a}})}\BibitemShut {NoStop}%
\bibitem [{\citenamefont {Imambekov}\ and\ \citenamefont
	{Demler}(2006{\natexlab{b}})}]{imambekov_applications_2006}%
\BibitemOpen
\bibfield  {author} {\bibinfo {author} {\bibfnamefont {A.}~\bibnamefont
		{Imambekov}}\ and\ \bibinfo {author} {\bibfnamefont {E.}~\bibnamefont
		{Demler}},\ }\href {\doibase 10.1016/j.aop.2005.11.017} {\bibfield  {journal}
	{\bibinfo  {journal} {Annals of Physics}\ }\textbf {\bibinfo {volume}
		{321}},\ \bibinfo {pages} {2390} (\bibinfo {year}
	{2006}{\natexlab{b}})}\BibitemShut {NoStop}%
\bibitem [{\citenamefont {Pollet}\ \emph {et~al.}(2006)\citenamefont {Pollet},
	\citenamefont {Troyer}, \citenamefont {Van~Houcke},\ and\ \citenamefont
	{Rombouts}}]{pollet2006phase}%
\BibitemOpen
\bibfield  {author} {\bibinfo {author} {\bibfnamefont {L.}~\bibnamefont
		{Pollet}}, \bibinfo {author} {\bibfnamefont {M.}~\bibnamefont {Troyer}},
	\bibinfo {author} {\bibfnamefont {K.}~\bibnamefont {Van~Houcke}}, \ and\
	\bibinfo {author} {\bibfnamefont {S.~M.~A.}\ \bibnamefont {Rombouts}},\
}\href {\doibase 10.1103/PhysRevLett.96.190402} {\bibfield  {journal}
{\bibinfo  {journal} {Physical Review Letters}\ }\textbf {\bibinfo {volume}
	{96}},\ \bibinfo {pages} {190402} (\bibinfo {year} {2006})}\BibitemShut
{NoStop}%
\bibitem [{\citenamefont {Adhikari}\ and\ \citenamefont
	{Salasnich}(2007)}]{adhikari_one-dimensional_2007}%
\BibitemOpen
\bibfield  {author} {\bibinfo {author} {\bibfnamefont {S.~K.}\ \bibnamefont
		{Adhikari}}\ and\ \bibinfo {author} {\bibfnamefont {L.}~\bibnamefont
		{Salasnich}},\ }\href {\doibase 10.1103/PhysRevA.76.023612} {\bibfield
	{journal} {\bibinfo  {journal} {Physical Review A}\ }\textbf {\bibinfo
		{volume} {76}},\ \bibinfo {pages} {023612} (\bibinfo {year}
	{2007})}\BibitemShut {NoStop}%
\bibitem [{\citenamefont {Mathey}\ and\ \citenamefont
	{Wang}(2007)}]{mathey_phase_2007}%
\BibitemOpen
\bibfield  {author} {\bibinfo {author} {\bibfnamefont {L.}~\bibnamefont
		{Mathey}}\ and\ \bibinfo {author} {\bibfnamefont {D.-W.}\ \bibnamefont
		{Wang}},\ }\href {\doibase 10.1103/PhysRevA.75.013612} {\bibfield  {journal}
	{\bibinfo  {journal} {Physical Review A}\ }\textbf {\bibinfo {volume} {75}},\
	\bibinfo {pages} {013612} (\bibinfo {year} {2007})}\BibitemShut {NoStop}%
\bibitem [{\citenamefont {Sengupta}\ and\ \citenamefont
	{Pryadko}(2007)}]{sengupta2007quantum}%
\BibitemOpen
\bibfield  {author} {\bibinfo {author} {\bibfnamefont {P.}~\bibnamefont
		{Sengupta}}\ and\ \bibinfo {author} {\bibfnamefont {L.~P.}\ \bibnamefont
		{Pryadko}},\ }\href {\doibase 10.1103/PhysRevB.75.132507} {\bibfield
	{journal} {\bibinfo  {journal} {Physical Review B}\ }\textbf {\bibinfo
		{volume} {75}},\ \bibinfo {pages} {132507} (\bibinfo {year}
	{2007})}\BibitemShut {NoStop}%
\bibitem [{\citenamefont {Rizzi}\ and\ \citenamefont
	{Imambekov}(2008)}]{rizzi_pairing_2008}%
\BibitemOpen
\bibfield  {author} {\bibinfo {author} {\bibfnamefont {M.}~\bibnamefont
		{Rizzi}}\ and\ \bibinfo {author} {\bibfnamefont {A.}~\bibnamefont
		{Imambekov}},\ }\href {\doibase 10.1103/PhysRevA.77.023621} {\bibfield
	{journal} {\bibinfo  {journal} {Physical Review A}\ }\textbf {\bibinfo
		{volume} {77}},\ \bibinfo {pages} {023621} (\bibinfo {year}
	{2008})}\BibitemShut {NoStop}%
\bibitem [{\citenamefont {Guan}\ \emph {et~al.}(2008)\citenamefont {Guan},
	\citenamefont {Batchelor},\ and\ \citenamefont {Lee}}]{guan2008magnetic}%
\BibitemOpen
\bibfield  {author} {\bibinfo {author} {\bibfnamefont {X.-W.}\ \bibnamefont
		{Guan}}, \bibinfo {author} {\bibfnamefont {M.~T.}\ \bibnamefont {Batchelor}},
	\ and\ \bibinfo {author} {\bibfnamefont {J.-Y.}\ \bibnamefont {Lee}},\ }\href
{\doibase 10.1103/PhysRevA.78.023621} {\bibfield  {journal} {\bibinfo
		{journal} {Physical Review A}\ }\textbf {\bibinfo {volume} {78}},\ \bibinfo
	{pages} {023621} (\bibinfo {year} {2008})}\BibitemShut {NoStop}%
\bibitem [{\citenamefont {Fang}\ \emph {et~al.}(2011)\citenamefont {Fang},
	\citenamefont {Vignolo}, \citenamefont {Gattobigio}, \citenamefont
	{Miniatura},\ and\ \citenamefont {Minguzzi}}]{fang2011exact}%
\BibitemOpen
\bibfield  {author} {\bibinfo {author} {\bibfnamefont {B.}~\bibnamefont
		{Fang}}, \bibinfo {author} {\bibfnamefont {P.}~\bibnamefont {Vignolo}},
	\bibinfo {author} {\bibfnamefont {M.}~\bibnamefont {Gattobigio}}, \bibinfo
	{author} {\bibfnamefont {C.}~\bibnamefont {Miniatura}}, \ and\ \bibinfo
	{author} {\bibfnamefont {A.}~\bibnamefont {Minguzzi}},\ }\href {\doibase
	10.1103/PhysRevA.84.023626} {\bibfield  {journal} {\bibinfo  {journal}
		{Physical Review A}\ }\textbf {\bibinfo {volume} {84}},\ \bibinfo {pages}
	{023626} (\bibinfo {year} {2011})}\BibitemShut {NoStop}%
\bibitem [{\citenamefont {Yin}\ \emph {et~al.}(2012)\citenamefont {Yin},
	\citenamefont {Guan}, \citenamefont {Zhang},\ and\ \citenamefont
	{Chen}}]{yin2012quantum}%
\BibitemOpen
\bibfield  {author} {\bibinfo {author} {\bibfnamefont {X.}~\bibnamefont
		{Yin}}, \bibinfo {author} {\bibfnamefont {X.-W.}\ \bibnamefont {Guan}},
	\bibinfo {author} {\bibfnamefont {Y.}~\bibnamefont {Zhang}}, \ and\ \bibinfo
	{author} {\bibfnamefont {S.}~\bibnamefont {Chen}},\ }\href {\doibase
	10.1103/PhysRevA.85.013608} {\bibfield  {journal} {\bibinfo  {journal}
		{Physical Review A}\ }\textbf {\bibinfo {volume} {85}},\ \bibinfo {pages}
	{013608} (\bibinfo {year} {2012})}\BibitemShut {NoStop}%
\bibitem [{\citenamefont {Guan}\ \emph {et~al.}(2013)\citenamefont {Guan},
	\citenamefont {Batchelor},\ and\ \citenamefont {Lee}}]{guan2013fermi}%
\BibitemOpen
\bibfield  {author} {\bibinfo {author} {\bibfnamefont {X.-W.}\ \bibnamefont
		{Guan}}, \bibinfo {author} {\bibfnamefont {M.~T.}\ \bibnamefont {Batchelor}},
	\ and\ \bibinfo {author} {\bibfnamefont {C.}~\bibnamefont {Lee}},\ }\href
{\doibase 10.1103/RevModPhys.85.1633} {\bibfield  {journal} {\bibinfo
		{journal} {Reviews of Modern Physics}\ }\textbf {\bibinfo {volume} {85}},\
	\bibinfo {pages} {1633} (\bibinfo {year} {2013})}\BibitemShut {NoStop}%
\bibitem [{\citenamefont {Hu}\ \emph {et~al.}(2016{\natexlab{a}})\citenamefont
	{Hu}, \citenamefont {Guan},\ and\ \citenamefont {Chen}}]{hu2016strongly}%
\BibitemOpen
\bibfield  {author} {\bibinfo {author} {\bibfnamefont {H.}~\bibnamefont
		{Hu}}, \bibinfo {author} {\bibfnamefont {L.}~\bibnamefont {Guan}}, \ and\
	\bibinfo {author} {\bibfnamefont {S.}~\bibnamefont {Chen}},\ }\href {\doibase
	10.1088/1367-2630/18/2/025009} {\bibfield  {journal} {\bibinfo  {journal}
		{New Journal of Physics}\ }\textbf {\bibinfo {volume} {18}},\ \bibinfo
	{pages} {025009} (\bibinfo {year} {2016}{\natexlab{a}})}\BibitemShut
{NoStop}%
\bibitem [{\citenamefont {Cazalilla}\ \emph {et~al.}(2011)\citenamefont
	{Cazalilla}, \citenamefont {Citro}, \citenamefont {Giamarchi}, \citenamefont
	{Orignac},\ and\ \citenamefont {Rigol}}]{cazalilla2011one}%
\BibitemOpen
\bibfield  {author} {\bibinfo {author} {\bibfnamefont {M.~A.}\ \bibnamefont
		{Cazalilla}}, \bibinfo {author} {\bibfnamefont {R.}~\bibnamefont {Citro}},
	\bibinfo {author} {\bibfnamefont {T.}~\bibnamefont {Giamarchi}}, \bibinfo
	{author} {\bibfnamefont {E.}~\bibnamefont {Orignac}}, \ and\ \bibinfo
	{author} {\bibfnamefont {M.}~\bibnamefont {Rigol}},\ }\href {\doibase
	10.1103/RevModPhys.83.1405} {\bibfield  {journal} {\bibinfo  {journal}
		{Reviews of Modern Physics}\ }\textbf {\bibinfo {volume} {83}},\ \bibinfo
	{pages} {1405} (\bibinfo {year} {2011})}\BibitemShut {NoStop}%
\bibitem [{\citenamefont {Imambekov}\ \emph {et~al.}(2012)\citenamefont
	{Imambekov}, \citenamefont {Schmidt},\ and\ \citenamefont
	{Glazman}}]{imambekov_one-dimensional_2012}%
\BibitemOpen
\bibfield  {author} {\bibinfo {author} {\bibfnamefont {A.}~\bibnamefont
		{Imambekov}}, \bibinfo {author} {\bibfnamefont {T.~L.}\ \bibnamefont
		{Schmidt}}, \ and\ \bibinfo {author} {\bibfnamefont {L.~I.}\ \bibnamefont
		{Glazman}},\ }\href {\doibase 10.1103/RevModPhys.84.1253} {\bibfield
	{journal} {\bibinfo  {journal} {Reviews of Modern Physics}\ }\textbf
	{\bibinfo {volume} {84}},\ \bibinfo {pages} {1253} (\bibinfo {year}
	{2012})}\BibitemShut {NoStop}%
\bibitem [{\citenamefont {Lin}\ \emph {et~al.}(2013)\citenamefont {Lin},
	\citenamefont {Matveev},\ and\ \citenamefont
	{Pustilnik}}]{lin_thermalization_2013}%
\BibitemOpen
\bibfield  {author} {\bibinfo {author} {\bibfnamefont {J.}~\bibnamefont
		{Lin}}, \bibinfo {author} {\bibfnamefont {K.~A.}\ \bibnamefont {Matveev}}, \
	and\ \bibinfo {author} {\bibfnamefont {M.}~\bibnamefont {Pustilnik}},\ }\href
{\doibase 10.1103/PhysRevLett.110.016401} {\bibfield  {journal} {\bibinfo
		{journal} {Physical Review Letters}\ }\textbf {\bibinfo {volume} {110}},\
	\bibinfo {pages} {016401} (\bibinfo {year} {2013})}\BibitemShut {NoStop}%
\bibitem [{\citenamefont {Ristivojevic}\ and\ \citenamefont
	{Matveev}(2016)}]{ristivojevic2016decay}%
\BibitemOpen
\bibfield  {author} {\bibinfo {author} {\bibfnamefont {Z.}~\bibnamefont
		{Ristivojevic}}\ and\ \bibinfo {author} {\bibfnamefont {K.~A.}\ \bibnamefont
		{Matveev}},\ }\href {\doibase 10.1103/PhysRevB.94.024506} {\bibfield
	{journal} {\bibinfo  {journal} {Physical Review B}\ }\textbf {\bibinfo
		{volume} {94}},\ \bibinfo {pages} {024506} (\bibinfo {year}
	{2016})}\BibitemShut {NoStop}%
\bibitem [{\citenamefont {Khodas}\ \emph {et~al.}(2007)\citenamefont {Khodas},
	\citenamefont {Pustilnik}, \citenamefont {Kamenev},\ and\ \citenamefont
	{Glazman}}]{khodas2007fermi-luttinger}%
\BibitemOpen
\bibfield  {author} {\bibinfo {author} {\bibfnamefont {M.}~\bibnamefont
		{Khodas}}, \bibinfo {author} {\bibfnamefont {M.}~\bibnamefont {Pustilnik}},
	\bibinfo {author} {\bibfnamefont {A.}~\bibnamefont {Kamenev}}, \ and\
	\bibinfo {author} {\bibfnamefont {L.~I.}\ \bibnamefont {Glazman}},\ }\href
{\doibase 10.1103/PhysRevB.76.155402} {\bibfield  {journal} {\bibinfo
		{journal} {Physical Review B}\ }\textbf {\bibinfo {volume} {76}},\ \bibinfo
	{pages} {155402} (\bibinfo {year} {2007})}\BibitemShut {NoStop}%
\bibitem [{\citenamefont {Gangardt}\ and\ \citenamefont
	{Kamenev}(2010)}]{gangardt_quantum_2010}%
\BibitemOpen
\bibfield  {author} {\bibinfo {author} {\bibfnamefont {D.~M.}\ \bibnamefont
		{Gangardt}}\ and\ \bibinfo {author} {\bibfnamefont {A.}~\bibnamefont
		{Kamenev}},\ }\href {\doibase 10.1103/PhysRevLett.104.190402} {\bibfield
	{journal} {\bibinfo  {journal} {Physical Review Letters}\ }\textbf {\bibinfo
		{volume} {104}},\ \bibinfo {pages} {190402} (\bibinfo {year}
	{2010})}\BibitemShut {NoStop}%
\bibitem [{\citenamefont {Tan}\ \emph {et~al.}(2010)\citenamefont {Tan},
	\citenamefont {Pustilnik},\ and\ \citenamefont
	{Glazman}}]{tan_relaxation_2010}%
\BibitemOpen
\bibfield  {author} {\bibinfo {author} {\bibfnamefont {S.}~\bibnamefont
		{Tan}}, \bibinfo {author} {\bibfnamefont {M.}~\bibnamefont {Pustilnik}}, \
	and\ \bibinfo {author} {\bibfnamefont {L.~I.}\ \bibnamefont {Glazman}},\
}\href {\doibase 10.1103/PhysRevLett.105.090404} {\bibfield  {journal}
{\bibinfo  {journal} {Physical Review Letters}\ }\textbf {\bibinfo {volume}
	{105}},\ \bibinfo {pages} {090404} (\bibinfo {year} {2010})}\BibitemShut
{NoStop}%
\bibitem [{\citenamefont {Karzig}\ \emph {et~al.}(2010)\citenamefont {Karzig},
	\citenamefont {Glazman},\ and\ \citenamefont {von
		Oppen}}]{karzig_energy_2010}%
\BibitemOpen
\bibfield  {author} {\bibinfo {author} {\bibfnamefont {T.}~\bibnamefont
		{Karzig}}, \bibinfo {author} {\bibfnamefont {L.~I.}\ \bibnamefont {Glazman}},
	\ and\ \bibinfo {author} {\bibfnamefont {F.}~\bibnamefont {von Oppen}},\
}\href {\doibase 10.1103/PhysRevLett.105.226407} {\bibfield  {journal}
{\bibinfo  {journal} {Physical Review Letters}\ }\textbf {\bibinfo {volume}
	{105}},\ \bibinfo {pages} {226407} (\bibinfo {year} {2010})}\BibitemShut
{NoStop}%
\bibitem [{\citenamefont {Micklitz}\ and\ \citenamefont
	{Levchenko}(2011)}]{micklitz2011thermalization}%
\BibitemOpen
\bibfield  {author} {\bibinfo {author} {\bibfnamefont {T.}~\bibnamefont
		{Micklitz}}\ and\ \bibinfo {author} {\bibfnamefont {A.}~\bibnamefont
		{Levchenko}},\ }\href {\doibase 10.1103/PhysRevLett.106.196402} {\bibfield
	{journal} {\bibinfo  {journal} {Physical Review Letters}\ }\textbf {\bibinfo
		{volume} {106}},\ \bibinfo {pages} {196402} (\bibinfo {year}
	{2011})}\BibitemShut {NoStop}%
\bibitem [{\citenamefont {Ristivojevic}\ and\ \citenamefont
	{Matveev}(2013)}]{ristivojevic2013relaxation}%
\BibitemOpen
\bibfield  {author} {\bibinfo {author} {\bibfnamefont {Z.}~\bibnamefont
		{Ristivojevic}}\ and\ \bibinfo {author} {\bibfnamefont {K.~A.}\ \bibnamefont
		{Matveev}},\ }\href {\doibase 10.1103/PhysRevB.87.165108} {\bibfield
	{journal} {\bibinfo  {journal} {Physical Review B}\ }\textbf {\bibinfo
		{volume} {87}},\ \bibinfo {pages} {165108} (\bibinfo {year}
	{2013})}\BibitemShut {NoStop}%
\bibitem [{\citenamefont {Matveev}\ and\ \citenamefont
	{Furusaki}(2013)}]{matveev_decay_2013}%
\BibitemOpen
\bibfield  {author} {\bibinfo {author} {\bibfnamefont {K.~A.}\ \bibnamefont
		{Matveev}}\ and\ \bibinfo {author} {\bibfnamefont {A.}~\bibnamefont
		{Furusaki}},\ }\href {\doibase 10.1103/PhysRevLett.111.256401} {\bibfield
	{journal} {\bibinfo  {journal} {Physical Review Letters}\ }\textbf {\bibinfo
		{volume} {111}},\ \bibinfo {pages} {256401} (\bibinfo {year}
	{2013})}\BibitemShut {NoStop}%
\bibitem [{\citenamefont {Ristivojevic}\ and\ \citenamefont
	{Matveev}(2014)}]{ristivojevic2014decay}%
\BibitemOpen
\bibfield  {author} {\bibinfo {author} {\bibfnamefont {Z.}~\bibnamefont
		{Ristivojevic}}\ and\ \bibinfo {author} {\bibfnamefont {K.~A.}\ \bibnamefont
		{Matveev}},\ }\href {\doibase 10.1103/PhysRevB.89.180507} {\bibfield
	{journal} {\bibinfo  {journal} {Physical Review B}\ }\textbf {\bibinfo
		{volume} {89}},\ \bibinfo {pages} {180507} (\bibinfo {year}
	{2014})}\BibitemShut {NoStop}%
\bibitem [{\citenamefont {Protopopov}\ \emph {et~al.}(2014)\citenamefont
	{Protopopov}, \citenamefont {Gutman},\ and\ \citenamefont
	{Mirlin}}]{protopopov2014relaxation}%
\BibitemOpen
\bibfield  {author} {\bibinfo {author} {\bibfnamefont {I.~V.}\ \bibnamefont
		{Protopopov}}, \bibinfo {author} {\bibfnamefont {D.~B.}\ \bibnamefont
		{Gutman}}, \ and\ \bibinfo {author} {\bibfnamefont {A.~D.}\ \bibnamefont
		{Mirlin}},\ }\href {\doibase 10.1103/PhysRevB.90.125113} {\bibfield
	{journal} {\bibinfo  {journal} {Physical Review B}\ }\textbf {\bibinfo
		{volume} {90}},\ \bibinfo {pages} {125113} (\bibinfo {year}
	{2014})}\BibitemShut {NoStop}%
\bibitem [{\citenamefont {Protopopov}\ \emph {et~al.}(2015)\citenamefont
	{Protopopov}, \citenamefont {Gutman},\ and\ \citenamefont
	{Mirlin}}]{protopopov_equilibration_2015}%
\BibitemOpen
\bibfield  {author} {\bibinfo {author} {\bibfnamefont {I.~V.}\ \bibnamefont
		{Protopopov}}, \bibinfo {author} {\bibfnamefont {D.~B.}\ \bibnamefont
		{Gutman}}, \ and\ \bibinfo {author} {\bibfnamefont {A.~D.}\ \bibnamefont
		{Mirlin}},\ }\href {\doibase 10.1103/PhysRevB.91.195110} {\bibfield
	{journal} {\bibinfo  {journal} {Physical Review B}\ }\textbf {\bibinfo
		{volume} {91}},\ \bibinfo {pages} {195110} (\bibinfo {year}
	{2015})}\BibitemShut {NoStop}%
\bibitem [{\citenamefont {Rozhkov}(2005)}]{rozhkov2005fermionic}%
\BibitemOpen
\bibfield  {author} {\bibinfo {author} {\bibfnamefont {A.~V.}\ \bibnamefont
		{Rozhkov}},\ }\href {\doibase 10.1140/epjb/e2005-00312-3} {\bibfield
	{journal} {\bibinfo  {journal} {The European Physical Journal B - Condensed
			Matter and Complex Systems}\ }\textbf {\bibinfo {volume} {47}},\ \bibinfo
	{pages} {193} (\bibinfo {year} {2005})}\BibitemShut {NoStop}%
\bibitem [{\citenamefont {Kulish}\ \emph {et~al.}(1976)\citenamefont {Kulish},
	\citenamefont {Manakov},\ and\ \citenamefont
	{Faddeev}}]{kulish1976comparison}%
\BibitemOpen
\bibfield  {author} {\bibinfo {author} {\bibfnamefont {P.~P.}\ \bibnamefont
		{Kulish}}, \bibinfo {author} {\bibfnamefont {S.~V.}\ \bibnamefont {Manakov}},
	\ and\ \bibinfo {author} {\bibfnamefont {L.~D.}\ \bibnamefont {Faddeev}},\
}\href {\doibase 10.1007/BF01028912} {\bibfield  {journal} {\bibinfo
	{journal} {Theoretical and Mathematical Physics}\ }\textbf {\bibinfo {volume}
	{28}},\ \bibinfo {pages} {615} (\bibinfo {year} {1976})}\BibitemShut
{NoStop}%
\bibitem [{\citenamefont {Orignac}\ \emph {et~al.}(2010)\citenamefont
	{Orignac}, \citenamefont {Tsuchiizu},\ and\ \citenamefont
	{Suzumura}}]{orignac_competition_2010}%
\BibitemOpen
\bibfield  {author} {\bibinfo {author} {\bibfnamefont {E.}~\bibnamefont
		{Orignac}}, \bibinfo {author} {\bibfnamefont {M.}~\bibnamefont {Tsuchiizu}},
	\ and\ \bibinfo {author} {\bibfnamefont {Y.}~\bibnamefont {Suzumura}},\
}\href {\doibase 10.1103/PhysRevA.81.053626} {\bibfield  {journal} {\bibinfo
	{journal} {Physical Review A}\ }\textbf {\bibinfo {volume} {81}},\ \bibinfo
{pages} {053626} (\bibinfo {year} {2010})}\BibitemShut {NoStop}%
\bibitem [{\citenamefont {Popov}(1972)}]{popov1972theory}%
\BibitemOpen
\bibfield  {author} {\bibinfo {author} {\bibfnamefont {V.~N.}\ \bibnamefont
		{Popov}},\ }\href {\doibase 10.1007/BF01028373} {\bibfield  {journal}
	{\bibinfo  {journal} {Theoretical and Mathematical Physics}\ }\textbf
	{\bibinfo {volume} {11}},\ \bibinfo {pages} {565} (\bibinfo {year}
	{1972})}\BibitemShut {NoStop}%
\bibitem [{\citenamefont {Haldane}(1981)}]{haldane_effective_1981}%
\BibitemOpen
\bibfield  {author} {\bibinfo {author} {\bibfnamefont {F.~D.~M.}\
		\bibnamefont {Haldane}},\ }\href {\doibase 10.1103/PhysRevLett.47.1840}
{\bibfield  {journal} {\bibinfo  {journal} {Physical Review Letters}\
	}\textbf {\bibinfo {volume} {47}},\ \bibinfo {pages} {1840} (\bibinfo {year}
	{1981})}\BibitemShut {NoStop}%
\bibitem [{\citenamefont {Lieb}\ and\ \citenamefont
	{Liniger}(1963)}]{lieb_exact_1963}%
\BibitemOpen
\bibfield  {author} {\bibinfo {author} {\bibfnamefont {E.~H.}\ \bibnamefont
		{Lieb}}\ and\ \bibinfo {author} {\bibfnamefont {W.}~\bibnamefont {Liniger}},\
}\href {\doibase 10.1103/PhysRev.130.1605} {\bibfield  {journal} {\bibinfo
	{journal} {Physical Review}\ }\textbf {\bibinfo {volume} {130}},\ \bibinfo
{pages} {1605} (\bibinfo {year} {1963})}\BibitemShut {NoStop}%
%\bibitem [{\citenamefont {Hu}\ \emph {et~al.}(2016{\natexlab{b}})\citenamefont
%	{Hu}, \citenamefont {Guan},\ and\ \citenamefont {Chen}}]{hu_strongly_2016}%
%\BibitemOpen
%\bibfield  {author} {\bibinfo {author} {\bibfnamefont {H.}~\bibnamefont
%		{Hu}}, \bibinfo {author} {\bibfnamefont {L.}~\bibnamefont {Guan}}, \ and\
%	\bibinfo {author} {\bibfnamefont {S.}~\bibnamefont {Chen}},\ }\href {\doibase
%	10.1088/1367-2630/18/2/025009} {\bibfield  {journal} {\bibinfo  {journal}
%		{New Journal of Physics}\ }\textbf {\bibinfo {volume} {18}},\ \bibinfo
%	{pages} {025009} (\bibinfo {year} {2016}{\natexlab{b}})}\BibitemShut
%{NoStop}%
\end{thebibliography}

%merlin.mbs apsrev4-1.bst 2010-07-25 4.21a (PWD, AO, DPC) hacked
%Control: key (0)
%Control: author (8) initials jnrlst
%Control: editor formatted (1) identically to author
%Control: production of article title (-1) disabled
%Control: page (0) single
%Control: year (1) truncated
%Control: production of eprint (0) enabled
%

\end{document}